\documentclass[12pt]{article}
\usepackage{multirow,verbatim}
\usepackage{latexsym, amsbsy, amssymb, geometry}
\usepackage{natbib}
\usepackage{lscape}
\renewcommand{\baselinestretch}{1.3}

\usepackage{tikz}
\usepackage{graphicx}
\usepackage{epsfig,float}
\usepackage{amsmath}
\usepackage{mathrsfs}
\usepackage{array}
\usepackage{enumerate} 
\usepackage[shortlabels]{enumitem}
\usepackage{multirow}
\usepackage{latexsym, amsbsy, amssymb}
\usepackage{verbatim}
\usepackage{color}
\usepackage{url}
\usepackage{caption}
\usepackage{subcaption}
\def\boe{\begin{enumerate}}
\def\eoe{\end{enumerate}}

\newtheorem{proposition}{{\bf Proposition}}
\newtheorem{lemma}{{\bf Lemma}}
\newtheorem{corollary}{{\bf Corollary}}

\newtheorem{theorem}{{\bf Theorem}}

\newtheorem{assump}{{\bf Assumption}}

\newcommand\ca[1]{{\cal{#1}}}
\newcommand\lo[1]{_{\nano{#1}}}
\newcommand\hi[1]{^{\nano{#1}}}

\def\E{\mathbb{E}}

\def\F{{\cal F}}

\def\L{{\cal L}}

\def\inv{^{\mbox{\tiny $-1$}}}

\newcommand{\indep}{\;\, \rule[0em]{.03em}{.65em} \hspace{-.41em}
\rule[-.02em]{.65em}{.03em} \hspace{-.41em}
\rule[0em]{.03em}{.65em}\;\,}

\def\ali{&\,}

\def\nano{\scriptscriptstyle}

\def\real{{\mathbb R}}

\def\ka{\kappa}

\def\L2T{L \lo 2 (T)}
\def\L2TX{L \lo 2 (T\lo X)}
\def\L2TX{L \lo 2 (T\lo Y)}

\def\ali{&\,}

\def\ali{&\,}

\def\eod{\end{document}}

\newfont{\rsfsten}{rsfs10 scaled 1100}
\newfont{\rsfstena}{rsfs10 scaled 800}
\newfont{\rsfstenb}{rsfs10 scaled 500}

\DeclareMathOperator*{\argmin}{argmin}
\DeclareMathOperator*{\essinf}{essinf}

\def\E{\mathbb{E}}

\newcommand{\ben}{\begin{enumerate}}
	\newcommand{\een}{\end{enumerate}}
\usepackage{xr}

\newcommand{\gamfull}[1][t,x]{\gamma(#1)}

\newcommand{\tgamfull}[2][t,x]{\tilde{\gamma}_{#2}(#1)}
\newcommand{\hgamfull}[2][t,x]{\hat{\gamma}_{#2}(#1)}

\newcommand{\hpropfull}[2][t,x]{\hat{f}_{#2}(#1)}

\newcommand\independent{\protect\mathpalette{\protect\independenT}{\perp}}
\def\independenT#1#2{\mathrel{\rlap{$#1#2$}\mkern2mu{#1#2}}}

\def\nano{\scriptscriptstyle}
\def\real{\mathbb R}

\def\Var{\mathrm{Var}}

\def\nano{\scriptscriptstyle}
\def\inv{\hi{\nano -1}}

\def\nano{\scriptscriptstyle}

\def\ka{\kappa}

\def\ali{&\,}

\newcommand{\RN}[1]{%
	\textup{\uppercase\expandafter{\romannumeral#1}}%
}
\newcommand{\mugm}{$\mu$g/m$^3$}
\newcommand{\mum}{$\mu$m}

\def \pmfine\  {PM$_{2.5}$}

\begin{document}

	\begin{center}
		{\Large{\bf
			Doubly robust estimation of causal effects for random object outcomes with continuous treatments}}

			\vskip.5cm
			{\large Satarupa Bhattacharjee$^\dagger$, Bing Li$^\ast$,  Xiao Wu$^\ddagger$, and Lingzhou Xue$^\ast$}\\
		\vskip.3cm
			$^\dagger$Department of Statistics, University of Florida\\ $^\ast$Department of Statistics, The Pennsylvania State University\\
			$^\ddagger$Department of Biostatistics, Mailman School of Public Health, Columbia University\\
		
   	\end{center}
	
	\begin{abstract}
Causal inference is central to statistics and scientific discovery, enabling researchers to identify cause-and-effect relationships beyond associations. While traditionally studied within Euclidean spaces, contemporary applications increasingly involve complex, non-Euclidean data structures that reside in abstract metric spaces, known as random objects, such as images, shapes, networks, and distributions. This paper introduces a novel framework for causal inference with continuous treatments applied to non-Euclidean data. To address the challenges posed by the lack of linear structures, we leverage Hilbert space embeddings of the metric spaces to facilitate Fr\'echet mean estimation and causal effect mapping. Motivated by a study on the impact of exposure to fine particulate matter ($\leq 2.5$\mum\ in diameter) on age-at-death distributions across U.S. counties, we propose a nonparametric, doubly-debiased causal inference approach for outcomes as random objects with continuous treatments. Our framework can accommodate moderately high-dimensional vector-valued confounders and derive efficient influence functions for estimation to ensure both robustness and interpretability. We establish rigorous asymptotic properties of the cross-fitted estimators and employ conformal inference techniques for counterfactual outcome prediction. Validated through numerical experiments and applied to real-world environmental data, our framework extends causal inference methodologies to complex data structures, broadening its applicability across scientific disciplines.
	\end{abstract}

\textbf{Keywords:} 
 Causal inference, Continuous treatment, Fr{\'e}chet regression, Semiparametric efficiency,  Air pollution and mortality.

\section{Introduction}
\label{sec:intro}

Causal inference is pivotal in statistics and scientific research, enabling the identification and estimation of cause-and-effect relationships beyond associations (e.g., \cite{rubi:74, rubi:05, holland1986statistics, pearl:16}). Going beyond randomized controlled trials~\citep{coln:24}, causal inference provides a powerful toolkit to analyze observational data and infer causal relationships, even in the presence of confounding factors, biases, and noise. By uncovering the underlying mechanisms driving observed phenomena, causal inference approaches facilitate better decision-making and more effective interventions across a broad range of disciplines.

However, most existing work in causal inference has focused on investigating causal effects within linear spaces, particularly within the Euclidean space $\mathbb{R}^p$. In contrast, the rise of complex non-Euclidean data, taking values in general metric spaces that often lack inherent linear structures, has become increasingly prominent in real-world applications.
Instances of such data, referred to as ``random objects'', include diverse forms such as images, shapes, networks, or life tables~\citep{marr:14}. Other notable examples include symmetric positive definite matrices,
networks, spherical surface data,~
and Riemannian manifolds, 
  among others. Given the metric space nature of the data, conventional statistical concepts such as sample or population means, defined as averages or expected values, do not readily apply and necessitate substitution with notions like barycenters or Fr\'echet means~\citep{frec:48}. In many modern applications, observed data either inherently manifests as or can be abstracted into such complex, non-Euclidean random objects. Often, the primary interest lies in understanding the causal effect on the random objects themselves. Consequently, there is a growing recognition that such applications are suitably characterized using non-Euclidean random objects. Modeling these as metric space-valued stochastic processes preserves their shape and geometry, providing richer information than scalar or vector summaries and necessitating new approaches to causal inference. 

A specific application that motivated our work is an environmental study, in which we examine the causal relation between the age-specific mortality distribution and the annual exposure to fine particles (with an aerodynamic diameter of 2.5\mum\ or smaller), denoted as \pmfine\  \ across the U.S.. The National Ambient Air Quality Standards (NAAQS) for \pmfine\  \  set the current primary standard for annual average at $9$ \mugm. While a body of literature concluded that exposure to \pmfine\  \  increases the risk of premature death among older adults~\citep{wu:20, jose:23}, the aggregate mortality rate is a scalar random variable that often fails to capture the age-specific mortality of the given region. Our interest is to summarize how the distribution of age-at-death,  which is a metric-space-valued random element, can be causally explained by continuously distributed \pmfine\  \ exposure in the presence of confounders while developing theoretical guarantees for the proposed estimator. This presents a significant challenge since the space of distribution-valued random variables lacks an inherent linear structure; as such, basic algebraic operations such as addition or scalar multiplication are not well defined in the space of distributions. However, one can consider the space of distributions, represented as quantile functions, CDF, or density functions, to be a metric space equipped with an appropriate metric such as the Wasserstein or Fisher-Rao metric~\citep{deli:17, le:17, pana:19}. 

In this paper, we develop an inverse probability weighted and a double-debiased estimation approaches for causal effects with non-Euclidean outcomes in the presence of moderately high-dimensional vector-valued confounding variables. We refer to this problem as CTROCIN (read as C-trocin), which stands for Continuous Treatment, Random Object Causal Inference. The regime of binary treatment and random object outcomes, referred to as BTROCIN, has been studied very recently by \citep{lin:23, kuri:24}. In the case of BTROCIN,~\cite{lin:23} developed a doubly robust estimation approach, which is limited to distributional outcomes, while~\cite{kuri:24} considered a specific, but often restrictive, modeling framework for outcomes in geodesic spaces. However, the general methodologies and theoretical foundations for  CTROCIN  remain underdeveloped. Extending from binary to continuous treatments introduces additional complexities, particularly in interpreting treatment effects across varying levels, due to potential selection bias. To address this challenge, we use dose density weights to estimate average causal responses. 
We propose a nonparametric doubly-debiased inference approach for non-Euclidean outcomes in general metric spaces that allow for embedding into some underlying Hilbert Spaces under the assumption of unconfoundedness given observed covariates.

The main contributions of this paper are as follows:
\begin{itemize}
\item[1.] Nonparametric CTROCIN: We introduce a novel nonparametric framework in Section 2 and a cross-fitted estimator for inferring causal effects in Subsection 3.3, when the treatment variable is continuous and the outcomes are metric-space-valued random objects embeddable in a latent Hilbert space. We give an in-depth discussion on the Hilbert space embedding property of a metric space and provide an efficient and useful way of computing Fr\'echet mean for random objects that allow for an injective map into a Hilbert space via a generalization of the Hahn-Banach separation theorem.
\item[2.] Model-Free Estimation via Efficient Influence Functions: Our approach, presented in Section 3, is nonparametric and data-driven, avoiding reliance on parametric models or distributional assumptions. We derive the moment function for the proposed estimate from a novel efficient influence function that applies to a wide class of parameters.
\item[3.] Rigorous Inference Framework: We develop a consistent and rigorous notational framework for continuous-treatment causal inference with random object outcomes. In Section 4, we discuss conformal predictive inference for important quantities of interest, which are suitably defined in the context of random object data. These include exact and asymptotic predictive inference for counterfactual outcomes, Fr\'echet exposure-dose function, Fr\'echet causal effect map, individual treatment effects, and so on. 
\item[4.] Asymptotic Theory: Section 4 establishes the asymptotic properties of the cross-fitted estimators, providing guarantees that ensure statistical validity and practical reliability. Numerical experiments in Section 5 further validate our theoretical findings.
\item[5.] Real-World Application: Section 6 demonstrates the utility of our methods by analyzing the causal effect of exposure to \pmfine\  \  on age-specific mortality distributions across U.S. counties, accounting for demographic, social, and environmental confounders. 
\end{itemize}

Section 7 includes a few concluding remarks. The additional numerical results, technical details, and complete proofs are presented in the supplement.

\section{Basic elements of CTROCIN}
\label{sec:prelim}
We consider the continuous evolution of the random object outcome $Y$ in response to a continuously varying treatment $T$, in the presence of possibly high-dimensional confounders $X$, in observational studies. 
The key ingredients of CTROCIN are summarized as follows: $(\Omega, \ca F, P)$, the probability space; $(\ca  T, \ca F \lo T)$, the treatment space, with  $T: \Omega \to \ca  T\subset \real$ being the treatment; $(\ca  X, \ca F \lo X)$, the covariate space, with $X: \Omega \to \ca  X\subset \real^p$ being the covariate; $(\ca  Y, d\lo Y)$, the metric space for the cross-sectional outcome $Y \lo t$ under treatment $t \in \ca  T$; $\ca M \lo Y$, the space of functions $\ca  T \to \ca  Y$, with  $Y: \Omega \to \ca M \lo Y$ being the outcome function, denoted by $Y =\{ Y\lo t: t\in \ca T \}$. We use $Y_t$ to represent the random-object-valued response at treatment level $t$. The random elements $X,T,Y$ are measurable with respect to (w.r.t.) $\ca F/ \ca F \lo X$, $\ca F/ \ca F \lo T$, and $\ca F / \ca F \lo Y$, respectively, where $\ca F \lo X, \ca F\lo T,$ and $\ca\F\lo Y$ are the Borel $\sigma$-fields on $\ca X,\ca T,$ and $\ca  Y$, induced by  their respective metrics. We denote 
$Z =(Y, T, X) \in \mathcal{Z}:= \mathcal{Y}\times \mathcal{T} \times \mathcal{X}$ from a population $\mathcal{P}$ with the CDF $F_Z$. By construction, the Stable Unit Treatment Values Assumption (SUTVA) holds: each subject’s potential outcome is unaffected by the treatment assignments of other subjects, and each treatment level is well-defined without hidden variations that could lead to different potential outcomes for the same unit.

\def\tfrac#1#2{\textstyle{\frac{#1}{#2}}}

The potential outcome function $Y$ in a causal setting is a function of $T$, $X$, and random unobserved heterogeneity. For Euclidean outcomes, a quantity of interest is often a summary measure of the potential outcome distribution that reflects the change due to the continuous treatment or exposure under the assumption of unconfoundedness~\ref{ass:unconfoundedness}.  A widely used measure is the \emph{average dose response} or exposure-response-function (ERF) defined as $E(Y\lo t)$, where  $Y\lo t$ denotes the potential outcome under the hypothetical treatment value $T=t$, and the expectation is taken over the distributions of $(X,\epsilon)$, with $\epsilon$ being the unobserved noise. When the outcome is a metric space-valued random object, however, linear functionals such as expectation or the usual additive error structure are not well-defined. Thus, understanding the effect of continuous treatment on random object responses requires leveraging the underlying geometry of the metric space. As such, the definition of an average or expected value is replaced by barycenters or Fr\'echet means~\citep{frec:48}. For any random variable $U$ in a metric space $(\ca M, d)$, its Fr\'echet mean is defined as $E\lo \oplus(U) := \argmin \{ E[ d\hi 2(U, u)] : u \in (\ca M, d)\}.$ 
Accordingly, for any potential outcome $Y\lo t$ at treatment level $t\in \ca T$ that takes value in a metric space $(\ca  Y,d\lo Y)$, the central tendency, interpreted as the `expected' potential distribution, is defined as the Fr\'echet mean
\begin{align}
	\label{fr:mean}
	\beta_t =  \E_\oplus(Y\lo t) := \argmin\lo{y\in \ca  Y} \E[d\lo Y\hi 2(Y\lo t,y)].
\end{align}
Henceforward, we will call this the \emph{Fr\'echet exposure-dose function (FERF)} 

A central challenge in causal effect estimation is that we do not observe $(X, T, \{ Y \lo t: t \in \ca  T\})$ for each subject. Instead, we observe only  $(X, T, Y \lo T)$ for each subject, where $Y \lo T$ is the cross-sectional response at the observed treatment $T$. It is helpful to view $Y\lo T$ as $\int Y\lo t d\delta_T(t)$, where $\delta_a$ is a Dirac measure at $a\in\ca T$, satisfying $\delta_a(B) = 1$ if $a\in B$  and $\delta_a(B) = 0$ otherwise. From the property of the Dirac measure that $\int f(t) d\delta_a(t) = f(a)$, the expression for $Y\lo T = \int Y \delta_T$ follows. This formulation separates the random outcome function $Y$ from the random treatment variable $T$. In particular,   this facilitates the derivation of the influence function and its semiparametric efficiency in Section 3 before we describe our proposed doubly robust estimator. 
We refer to the map $(X, T, Y) \mapsto (X, T, Y\lo T) = (X, T, \int\lo{\ca T}Y \delta_T )$ as the observation mapping, since only the image of this map is observed.

In general, for any given $t \in \ca  T$, $E(Y \lo T|T = t)$ does not give us an unbiased estimate of $E(Y \lo t)$, because $T$ may be affected by other factors, known as confounders, that also affect $T$ and $Y$. To account for confounding, we introduce the covariate $X$, which we assume contains all the confounders: after conditioning on $X$, the response $Y$ no longer depends on $T$. This assumption, known as ignorability, conditional independence, or selection of observables, is  standard in the causal inference literature~\citep{rubi:74, rubi:05}:
  \begin{assump}
       \label{ass:unconfoundedness} \textbf{Ignorability}: $Y\independent T |X$
  \end{assump} 
This assumption asserts that, conditional on observables, the treatment assignment is conditionally exogenous or behaves as if it were randomized. It implies that the observational study is similar to a randomized controlled trial that facilitates valid estimation of causal effects. To achieve valid inference, we adopt a doubly debiased machine learning approach that leverages a doubly robust moment function combined with cross-fitting,  as long as the response space admits a suitable embedding into a Hilbert space. Our approach is fully nonparametric,  imposing no distributional or functional form assumptions on $T$, $X$, or $\epsilon$.

\subsection{Embedding of a metric space into a Hilbert space}
Embedding metric spaces into simpler and more structured spaces that have low distortion plays an important role in the analysis of random objects, and such embeddings have widespread applications across fields. Proposition 3 of~\cite{sejd:12} implies that whenever $d\lo Y$ is a semi-metric of
negative type, there exists a Hilbert space $\ca H$ and an injective map, say $f : \ca  Y \to \ca H$ with $d\lo Y\hi 2(Y_1,Y_2) = \|f(Y_1) - f(Y_2)\|\lo{\ca H}\hi 2,$ for any $Y_1, Y_2 \in \ca  Y$.
Thus, if the metric space $(\ca  Y,d\lo Y)$ where the outcome function takes values in is of strong negative type, the existence of an isometric continuous embedding from $\ca  Y $ to an underlying Hilbert space $\ca H$ is guaranteed. Here,  
a space $(M,\rho)$ with a semi-metric $\rho$ is of negative type if for all $n \geq 2$, $z_1, z_2, \dots,z_n \in M$ and $\alpha_1,\alpha_2,\dots,\alpha_n \in \real$, with $\sum_{i=1}^n \alpha\lo i = 0,$ one has $\sum_{i=1}^n\sum_{j=1}^n \alpha\lo i \alpha_j \rho(z\lo i,z_j) \leq 0$.
Every separable Hilbert space is of strong negative
type. An explicit form of such a continuous, injective, isometric map is given in the following theorem.
\begin{lemma}[~\cite{sejd:12}]
Suppose $(\ca  Y, d\lo Y)$ is a separable metric space.  Denote $\kappa(z, z') =
\frac{1}{2}[d\lo Y(z, z_0)+ d\lo Y(z', z_0)- d\lo Y(z, z')]$. Then $\kappa$ is positive definite if and only if $(\ca  Y, d\lo Y)$ is of negative type. Thus there exists an Aronszajn map $\rho: z\mapsto \kappa(\cdot,z)$, which is an isometric embedding of the metric space $(\ca  Y,\rho)$ into the RKHS, $\ca H \lo{\kappa}$, generated by $\kappa$, i.e., (1) $\rho(y) = \kappa(\cdot, y)$ is a continuous injection, and (2) $\lVert\rho(z_1) - \rho(z_2) \rVert\lo{\ca H} = d\lo Y(z_1,z_2).$
\end{lemma}
Next, we discuss special cases for a metric space to be embeddable in a Hilbert space. The constructions of the Hilbert space embeddings for commonly observed random object data, namely distributional objects, SPD matrices, compositional data, and phylogenetic trees, are presented in the supplement, with more detailed literature review and discussion along with examples. 

\noindent \textbf{Embedding for the space of probability distribution:} Let $(\ca P,d\lo Y)$ denote the space of probability distributions on a measurable space $(\ca X, \ca B\lo{X})$, and let $\kappa:\ca X\times \ca X\to \real$ be a measurable, positive definite kernel with associated RKHS $\ca H\lo{\kappa}$, such that $\sup\lo{x\in \ca X} \ka(x,x)<\infty$. The kernel mean embedding of the probability measure $P$ is defined as the Bochner integral of $\ka(\cdot,x)$ w.r.t. $P$, that is,
$\rho: P\mapsto \int \ka(\cdot,x) dP(x), \text{ for } P\in \ca P,$
and $\rho$ is a continuous injective map if the kernel is characteristic~\citep{fuku:04}, meaning that $\rho(P)$ uniquely represents $P\in \ca P$,  preserving all information. Examples of characteristic
kernels on $X = \real^d$ include Gaussian, Mat\'ern, and Laplace kernels~\citep{srip:10}. Note that the space of the probability distribution is convex; thus, for any $P,Q\in \ca P$ $\lambda P+ (1-\lambda)Q \in \ca P$ for $0<\lambda<1$. Combined with the linearity of the integral operation, this yields that $\rho(\ca P)$ is convex. Furthermore, for continuous and bounded kernel functions, $\rho(\ca P)$ is closed by the portmanteau lemma.

\noindent \textbf{Embedding for the space of SPD matrices and networks:} The cone of $K\times K$ symmetric positive semi-definite matrices, $\mathcal{S}_K$, equipped with a suitable choice of metric, such as the Frobenius metric, log-Euclidean metric~\citep{arsi:07}, the power metric family~\citep{dryd:10,pigo:14, tava:19}, the Log-Cholesky metric~\citep{lin:19},  the Bures-Wasserstein metric \citep{taka:11}, and so on induce a Riemannian manifold structure on $\mathcal{S}_K$~\citep{bhat:09}.

\begin{figure}[htp]
\centering
\includegraphics[width=.3\textwidth]{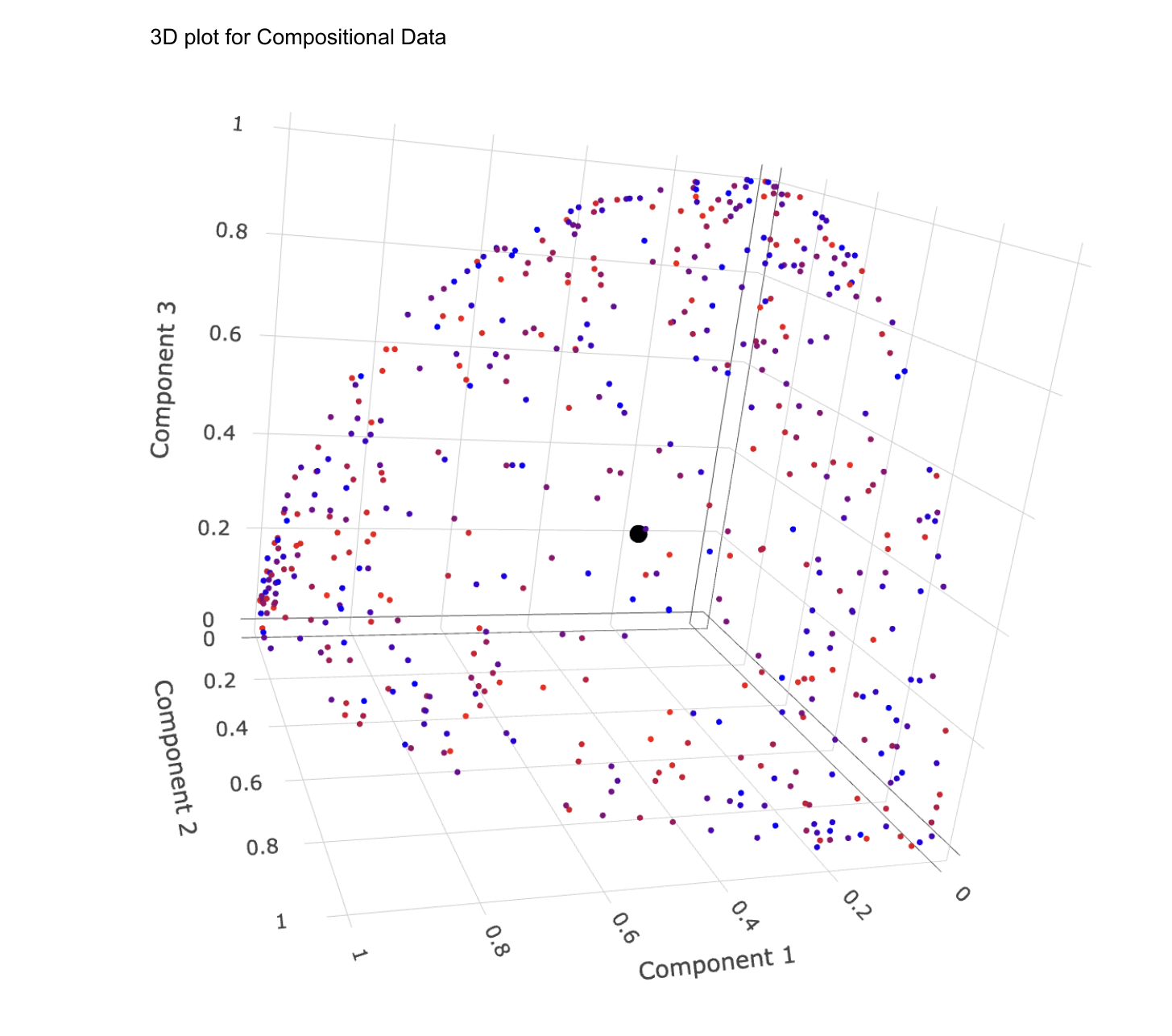}\hfill
\includegraphics[width=.3\textwidth]{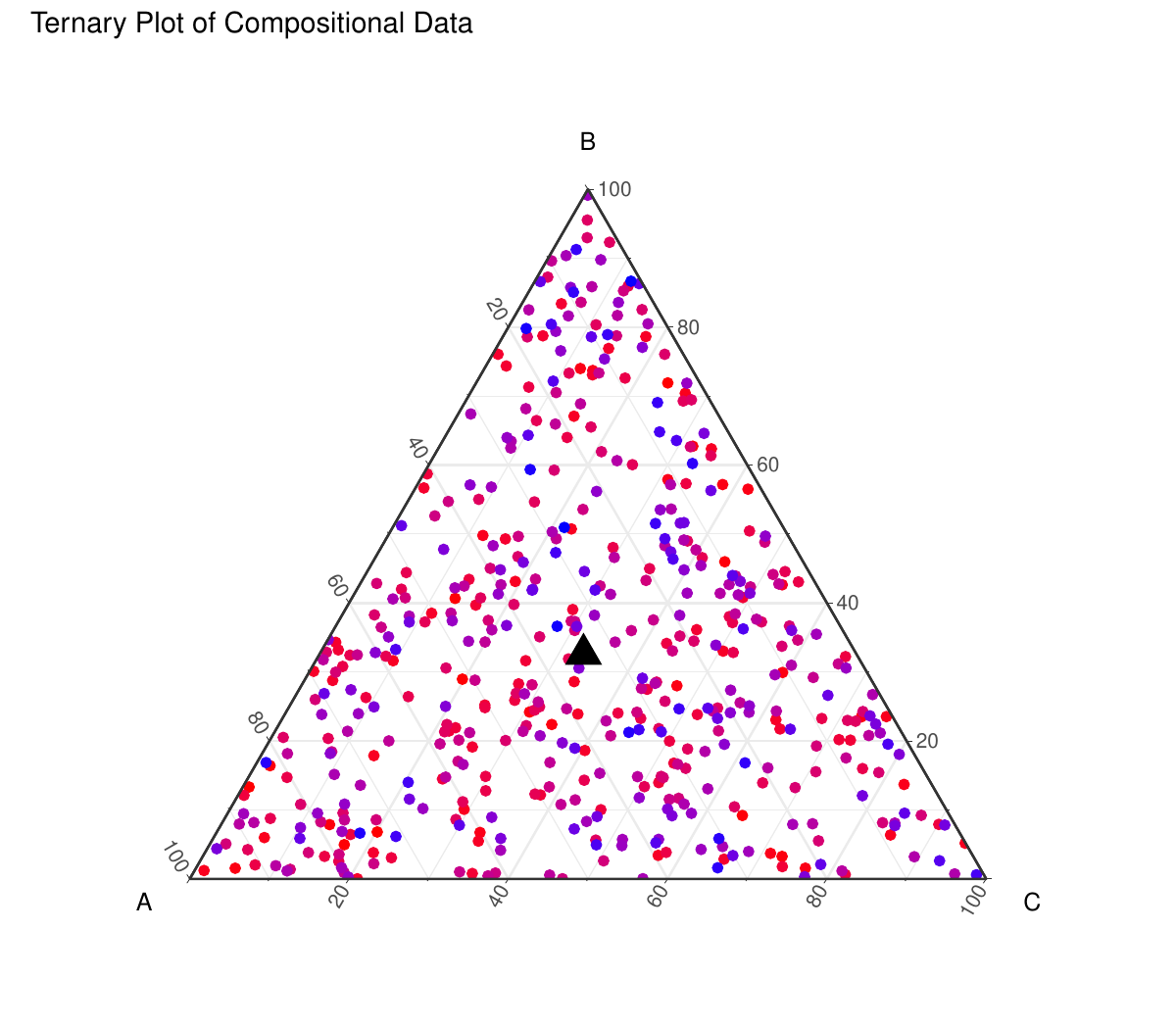}\hfill
\includegraphics[width=.3\textwidth]{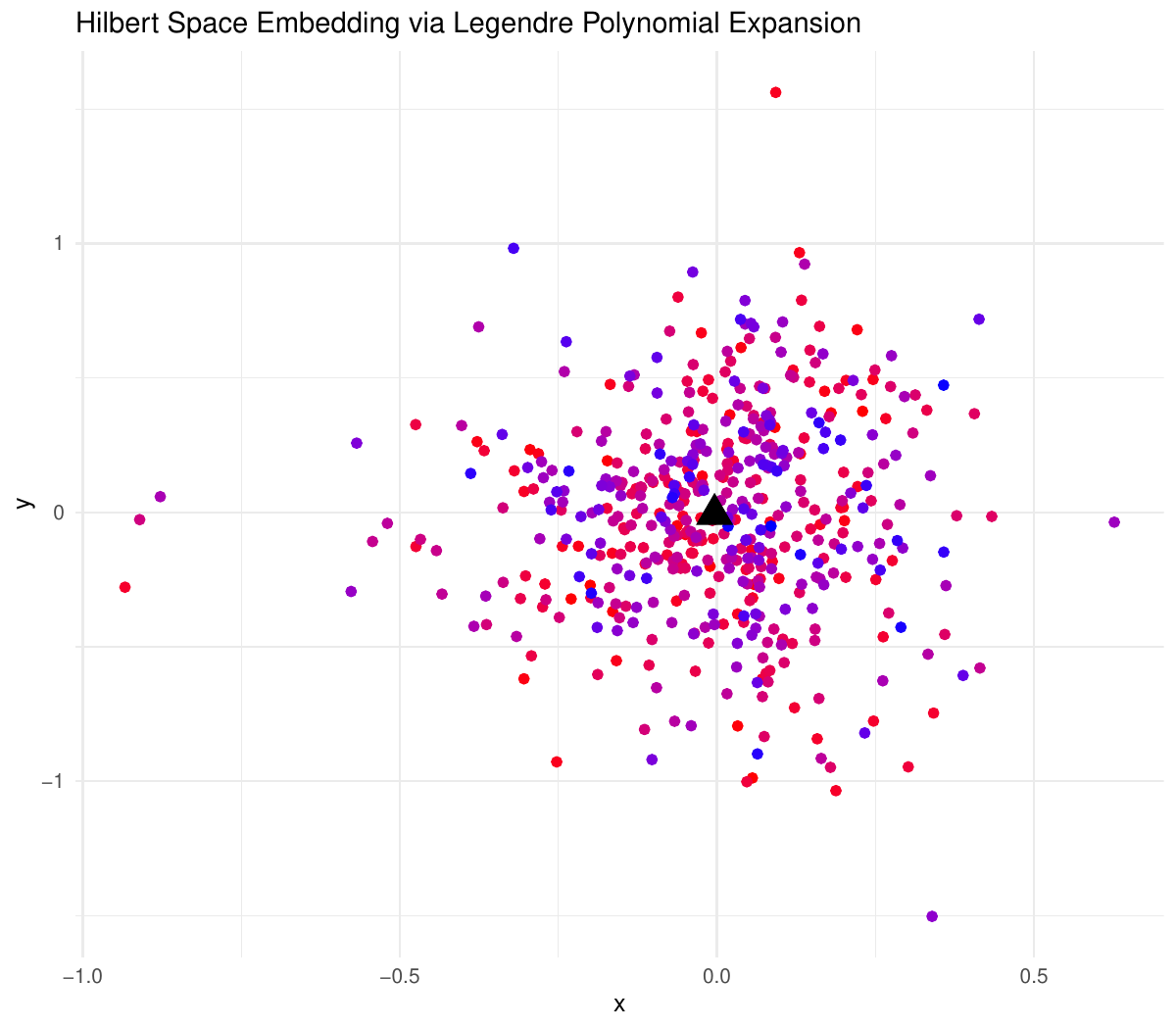}
\caption{Illustration of a toy example with $200$ points generated on the positive quadrant of the unit sphere $\mathcal{S}^2 \subset \real^3.$ The left panel shows a 3D scatter plot; the middle panel shows a ternary plot that represents the compositional data as simplices under Aitchison's geometry; and the right panel shows the Legendre polynomial embedding in $\real^3$. The Fr\'echet mean (w.r.t. the geodesic distance) is shown as a black triangular point in all panels. 
}
\label{fig:illustration:comp}
\end{figure}

\noindent \textbf{Embedding for the finite-dimensional Riemannian manifold:} In the context of Riemannian manifolds, mapping data into a Hilbert space is well-studied. Embedding into a Reproducing Kernel Hilbert Space (RKHS) can be achieved by using heat kernels~\citep{bera:94, chu:22} or Gaussian RBF kernels~\citep{jaya:15, jaya:16}, with appropriate adjustments for the curvature of the manifold. 
Figure~\ref{fig:illustration:comp} shows an illustration of a Hilbert space embedding for compositional data situated on the surface of a sphere, using a Legendre polynomial embedding map. 

\noindent \textbf{Embedding for Phylogenetic trees:} Tree space~\citep{bill:01} is an example that may not admit a Riemannian structure. Phylogenetic trees are widely used in evolutionary biology to represent the ancestral relationships among a set of organisms, and a vector space embedding of tree space is possible using ``tropical geometry''~(e.g., \cite{song:11}).

\subsection{Fr\'echet exposure-dose response function (FERF)}

The existence and uniqueness of the Fr\'echet means depend on the nature of the space, as well as the metric considered. For example, in the case of Euclidean responses, the Fr\'echet means coincide with the usual means for random vectors with finite second moments. In the case of Riemannian manifolds, the existence, uniqueness, and convexity of the center of mass are guaranteed~\citep{afsa:11, penn:18}. In a space with a negative or zero curvature, or a Hadamard space, unique Fr\'echet means are also shown to exist (see e.g.~\cite {stur:03}).

An efficient way to compute the Fr\'echet means is to embed the metric space into a Hilbert space, as discussed in the previous subsection, and use the Riesz representation theorem to compute the expectation within that space. This lends more structure to the abstract metric space, enabling the use of Hilbert space geometry, such as the inherent direction and interpretability of linear space, to support nonparametric or functional causal inference.

\begin{assump}
\label{ass:isometry}  There is a Hilbert space  $\ca H$ and a continuous injection  $\rho: \ca  Y  \to \ca H$ 
    such that  $\rho: \ca  Y  \to \rho (\ca  Y )$ is an isometry.
\end{assump}

\begin{assump}
\label{ass:convex} 
The set   $\rho(\ca  Y )$ is convex and closed in $\ca H$.
\end{assump}

The following result, which we rely on heavily in the subsequent development, seems not to have been recorded in the literature to the best of our knowledge. So we formally state it here and rigorously prove it in the Supplementary Material. For a random element $V$ taking values in a Hilbert space $\ca H$, we define its expectation as the Riesz representation of the linear functional that maps $f \in {\ca H}$ to $E \langle f, V \rangle \lo {\ca H} \in \real$. This linear functional is  bounded if $E \| V \| \lo {\ca H} < \infty$.

\begin{proposition}
If $C$ is a closed and convex subset of a Hilbert space $\ca H$ and $V$ is a random element taking values in $\ca H$ with $E \| V \| \lo {\ca H} < \infty$,  then   $E(V)$ defined above is a member of   $C$.
\end{proposition}

Hereafter, we assume the form of the continuous injective map $\rho$ is known. The object responses $Y\lo t$ are thus embedded in the Hilbert space $\ca H$, and the effective outcomes are denoted as $V\lo t=\rho(Y\lo t)$ for each $t\in \ca T$. 
Let $\ca M \lo V$ be a space of $\ca H$-valued functions defined on $\ca T$. Assume, for each $\omega \in \Omega$,  the function $V (\omega)$ defined by  $t \mapsto V \lo t(\omega)$ is a member of $\ca M \lo V$. Thus,  the mapping $V: \omega \mapsto V(\omega)$ from $\Omega$ to $\ca M \lo V$ defines a random element in $\ca M \lo V$. In this notation system, $\{\ca H, V, \ca M \lo V\}$ are counterparts of $\{\ca Y, Y, \ca M \lo Y\}$ after the Hilbert-space embedding.
 This  embedding $\rho$ allows us to compute $E \lo \oplus(Y \lo t)$ through operations in the Hilbert space. Under Assumptions~\ref{ass:isometry} and~\ref{ass:convex}, the FERF, defined as the Fr\'echet mean of the responses $Y\lo t \in (\ca  Y,d\lo Y)$ in~\eqref{fr:mean} can be equivalently written as 
\begin{align}
    \label{eq:ERF}
E \lo \oplus ( Y \lo t) = \rho \inv ( E (\rho (Y \lo t))). 
\end{align}
The significance of this relation is that $\rho (Y \lo t)$ is a Hilbert-space-valued random element, and its expectation can be computed by Riesz representation and thus can be estimated. In fact, we can rewrite the FERF estimand in a computationally cheap and interpretable way under the Hilbert space embedding as follows.  Our strategy is to estimate $E(V \lo t)$ in a Hilbert space and then transform the result by $\rho \inv$ to estimate $\beta\lo t= E \lo \oplus (Y \lo t)$. 
\begin{proposition}
\label{Prop:towering:Expectation}
    Under Assumptions~\ref{ass:isometry} and~\ref{ass:convex}, we have 
\begin{enumerate}
    \item $\rho \inv ( E (\rho (Y \lo t))) $ is well defined and $  E \lo \oplus (Y \lo t)   = \rho \inv ( E (\rho (Y \lo t))) $; 
    \item $\rho ( E \lo \oplus (Y \lo t) ) = E [ \rho ( E \lo \oplus ( Y \lo t | X)] $, where $E \lo \oplus (Y \lo t |X = x)$ is defined as the minimizer of $E[ d \hi 2  (Y \lo t, y)|X = x]$ among all  $y\in \ca Y$, which is interpreted as the outcome regression in the object space.
\end{enumerate}
\end{proposition}

In the following, we use $m(t, X)$ to denote $E \lo \oplus ( Y \lo t | X = x)$,  and call it the Fr\'echet conditional ERF. Let us denote $\gamfull:= \rho(m(t,x))$.
Further, the causal effects map between two different levels of treatments $t$ and $t'$ can be quantified as $\Delta\lo{t,t'}:= \E(\|V\lo t  - V\lo{t'}\|\lo{\ca H})$, again via the embedding map $\rho$.

\section{Estimation}
\label{sec:est}

In this section ,we introduce two estimators of the CITROCIN: one based on inverse probability weighting (IPW), and the other on doubly robust estimation (DR).  We assume that $\rho$ is a known bijective transformation satisfying Assumptions~\ref{ass:isometry}--\ref{ass:convex}. After the transformation, we have an i.i.d. sample $\{X_i, T_i, V_i\}_{i=1}^n$. As mentioned earlier, our strategy is to estimate $E(V \lo t)$ and then employ the relation $E \lo \oplus (Y \lo t) = \rho \inv ( E (V \lo t))$ to estimate $ E \lo \oplus (Y \lo t)$. Since we   observe $V \lo T$ but not $V \lo t$, our estimator must be based on $V \lo T$ instead of  $V \lo t$, and this is the main challenge in causal estimation.

\subsection{ Inverse probability weighting approach for CTROCIN}

This subsection constructs a causally unbiased estimate of $\rho(\beta\lo t) = E(V \lo t)$ by inverse probability weighting (IPW); see, for example, \cite{imbe:15} and \cite{pearl:16}. Since $T$ is continuous, the probability of observing $T = t$ is zero. Thus, combining information from nearby points is necessary.  We approach this by first approximating $\beta \lo t$ as follows. Let 
\begin{align}\label{eq:diffused EYt}
\vartheta \lo t ( h) =  \int \lo {\ca T}  E(V \lo s ) K \lo h ( s - t) d s,  
\end{align}
where the integral is taken as Brochner integral \citep{hsing2015theoretical}, $K \lo h (u) = k (u / h) / h$, $h$ is a positive constant, and $k$ is a probability density function defined on $\ca T$.
By construction, 
$\lim \lo {h \to 0}  \vartheta \lo t (h) = \beta \lo t. 
$
So, to construct a consistent estimate of $\beta \lo t$, we let $h\to 0$ as $n \to \infty$. 

Subsequently, we consider a kernel function $K_h$ satisfying the following integrability assumptions
\begin{assump}
\label{ass:kernel} \textbf{Kernel integrability conditions }: The kernel function $k: \real \to \real$ is symmetric satisfying $k(u) = k(-u)$ for $u \in \real$, and $\int_{\real} k(u) du =1$. Furthermore,  $\int_{-\infty}^{\infty} k(u) du = 1$, $\int_{-\infty}^\infty uk(u)du = 0,$ and $\int_{-\infty}^\infty u^2k(u)du<\infty$. Also, $|k'(u)|\leq C|u|^{-a}$ for $|u|>C_2$ for some finite positive constants $C_1,C_2$ and $a>1$.
\end{assump}

  For an $a \in \ca T$, let $\delta \lo a$ be the Dirac measure at $a$. Let $w: \ca T \to \real$ be an arbitrary nonnegative measurable function on $\ca T$. In particular, it could be the function $s \mapsto K \lo h ( s - t)$ for the fixed $t$ in $V \lo t$.  Our IPW estimator is based on the following population-level result. The key rule we have to obey when constructing a causal estimator of $E(V \lo t)$ is that we are only allowed to use $V \lo T$, as $V \lo t$ is not observed for any fixed $t \in \ca T$.  Note that, in terms of the Dirac measure $\delta \lo a$, $V \lo T$ can be rewritten as the integral form $\int \lo {\ca T} V \lo s  d \delta \lo T (s)$. The benefit of using this alternative expression is that it separates $V$ from $T$, allowing us to apply the conditional independence $Y \indep T |X$ in a more intuitive fashion. 
\begin{proposition}\label{proposition:ipw} If $V \lo t$ is Brochner integrable with respect to $P$,  $w(t) E( Y \lo t)$ is Brochner  integrable with respect to the Lebesgue measure on $\ca T$ and $f \lo {T|X}(t|x) > 0$ for any $t \in \ca T$ and $x \in \ca X$, then 
\begin{align*}
 E \left(   \int \lo {\ca T } \frac{w(t)} { f \lo {T|X} (t |X)} V \lo t d \, \delta \lo T (t) \right) = \int \lo {\ca T} w(t) E (V \lo t) dt. 
\end{align*}
\end{proposition}

The point of this equality is that the left-hand side depends on $V \lo T$ but not $V \lo t$; whereas the right-hand side does depend on $V \lo t$. The IPW estimator is based on the mimicry of the left-hand side of the above equality 
at the sample level. In fact, if we choose $w(t)$ to be the kernel function $K \lo h ( s- t)$, 
 then, by letting $h \to 0$, we can prove the following limit form of the above equality in Proposition~\ref{proposition:ipw}. The conditional density $f_{T|X}$, a.k.a. the General Propensity Score (GPS),  plays a central role.  %
The following assumption on the smoothness of the GPS is used in the subsequent sections whenever needed.
\begin{assump}
\label{ass:propensity}
\textbf{Generalized Propensity Score}: The propensity score $f\lo{T|X}$ is bounded away from $0$, i.e., there exists a positive constant $c$ such that $\inf\lo{t\in \ca T}\essinf\lo {x\in\ca X}f\lo{T|X}(t|x) \geq c.$
 Furthermore, we assume that $f\lo{T|X}(t|x)$ is a three-times differentiable function w.r.t. $t$ with $\sup_{x\in\ca X} \  \frac{d\hi k}{dt\hi k}f\lo{T|X}(t|x) \leq c_k$ for some positive constants $c_k$, $k=1,2,3.$ 
 \end{assump}

\begin{corollary} Suppose 
\begin{enumerate}
    \item $\ca T = (-\infty, \infty)$, and     
$k(\cdot)$ is a bounded probability density function on $(-\infty, \infty)$;  \vspace{-0.12in}
    \item there is a measurable function $g: \Omega \to \real$ such that $\int g (\omega) d P (\omega) < \infty$ and $\| V \lo t (\omega) \| \lo {\ca H} < g (\omega)$ for all $\omega \in \Omega$;  there exist $c \lo 1 > 0$ and $c \lo 2 < 0$ such that $f \lo T (s) < c \lo 2$ and $f \lo {T|X}(s|x) > c \lo 1$ for all $s \in \real$ and $x \in \ca X$; \vspace{-0.12in}
    \item for each $\omega \in \Omega$, the mapping $V \lo s (\omega)$ is continuous at $s = t$;  $f \lo T (s)$ and $f \lo {T|X} (s|x)$ are continouous at $t$; $E ( \| V \lo s \| \lo {\ca H} | T = s)$ is bounded and continuous at $s = t$.   
\end{enumerate}
Then \vspace*{-.2in}
\begin{align*}
    E (V \lo t ) = E \left[ \left.  \frac{f \lo T (t)}{f \lo {T|X} (t |X)} V \lo T \right| T = t \right]. 
\end{align*}

\end{corollary}

As mentioned earlier, we choose the weight function $w(t)$ in Proposition \ref{proposition:ipw} to be a kernel function $k ((s-t)/h)/h$ for some probability density function on $\ca T$.

\subsection{Doubly robust approach for  CTROCIN}

Inverse probability weighting described in the last subsection is essentially a moment estimator that allows re-expressing the moment $E(\int \lo {\ca T} w(t) V \lo t d t )$ in terms of the observable variable $V \lo T$. An alternative approach is the doubly robust estimator, which is semiparametrically efficient under regularity conditions. In this section, we develop such an estimate. Similar to the IPW case, we first target $\vartheta \lo t ( h) = E(\int \lo {\ca T} K \lo h (s - t)  V \lo s d s )$, and then let $h$ go to 0 to estimate $E (V \lo t)$. The next theorem gives the semiparametrically efficient influence function for estimating the real-valued parameter $E(\int \lo {\ca T} w(s)  V \lo s d s )$, where $w(s)$ is a general weighting function, and when the infinite-dimensional nuisance parameters -- $f \lo X$, $f \lo {T|X}$ and $\{f \lo {V \lo t |X}: t \in \ca T \}$ --- are  completely unknown.

\begin{theorem}\label{theorem:eff score}
   Suppose $V \indep T | X$  and $f \lo {T|X}(t|x) > 0$ for all $t \in \ca T$ and $x \in \ca X$. 
    Then, the efficient influence  function $\phi ( X, T, V \lo T)$ for estimating $E ( \int \lo {\ca T} w (s) V \lo s \, d s)$ is
 \begin{align}\label{eq:efficient score}
\phi (X, T, Y \lo T)   
= \int \lo {\ca T } \left(  \frac{w(s)  }{f \lo {T|X} (s | X) }  W (V \lo s, X)  \right) d \delta \lo T (s)+ B (X) .
\end{align}      
where  
$
B  = \int \lo {\ca T}w(s)   [ E( V \lo s | X) - E (V  \lo s ) ] d s  
$, and $\
W (V \lo s, X) =
V \lo s - E(V  \lo s |X). 
$   
Equivalently, the efficient score function can be re-expressed as 
\begin{align*}
 \phi (X, T, V \lo T  )   
 = \ali  \frac{ w(T) }{f \lo {T|X} (T | X) }  \left[ V \lo T 
  -  E (V \lo T  | X, T) \right]    +   E \left[ \left. \frac{w (T) }{f \lo {T|X} (T|X)}   V \lo T  \right| X  \right] 
 - E \left( \int \lo {\ca T} w(s) V \lo s \, d s  \right). 
\end{align*} 

In particular, 
the efficient influence  function $\phi ( X, T, V \lo T)$ for estimating $\vartheta \lo h ( t) = E ( \int \lo {\ca T} K \lo h (s-t) V \lo s \, d s)$ is
\begin{align}\label{eq:efficient score for vartheta}
 \phi (X, T, V \lo T  )   
 = \ali  \frac{K \lo h ( T- t) }{f \lo {T|X} (T | X) }  \left[ V \lo T 
  -  E (V \lo T  | X, T) \right]    +   E \left[ \left. \frac{K \lo h (T - t)}{f \lo {T|X} (T|X)}   V \lo T  \right| X  \right] 
 - \vartheta \lo t (h). 
\end{align} 
\end{theorem}  

In the binary treatment setting where $\ca T = \{0,1\}$, the potential outcome framework closely aligns with the missing at random (MAR) problem, allowing the average treatment effect (ATE) at two levels to be formulated as a semiparametric estimand. Here, ATE is the parametric component, while functions $f\lo X, f\lo{T|X}, f \lo {V \lo 0|X}, f \lo {V \lo 1 | X}$ are the nonparametric components or the infinite-dimensional nuisance parameters. This setting admits only one influence function in this scenario, so it is semiparametrically efficient. This theory easily extends to the case of finite treatment levels $k$, where standard semiparametric theory can be used to derive the efficient influence function, efficiency bound, and corresponding estimator.
However, extending this framework to the continuous treatment setting is nontrivial, as standard semiparametric theory does not apply directly. To address this gap, the key insight of our proof of Theorem \ref{theorem:eff score} is to introduce the random Dirac measure $\delta \lo T$ to separate the random function $V$ and the random vector $(X, T)$.

Since our goal is to estimate $\rho ( \beta \lo t) = \vartheta \lo t = E (V \lo t)$ instead of $\vartheta \lo t (h)$, we further derive below the limiting   (as $h \to 0$) moment condition derived from the efficient score (\ref{eq:efficient score for vartheta}).  

\begin{corollary}\label{corollary:limit form of phi} Suppose,  (1)
$f \lo {T|X} (t|x) > 0$ for all $t \in \ca T$ and $x \in \ca X$; and (2) the function $ s \mapsto   E \left(\left. {V \lo s}  \right| X = x, T = s   \right) $ is continuous at $t$ for every $x$ and its absolute value is bounded from above by a function $g(x)$ that is integrable with respect to $P \lo X$. 
Then 
\begin{align*}
   \vartheta \lo t  = \lim \lo {h\to 0} E \left( \frac{K \lo h (T - s)}{f \lo {T|X} (T | X )} \, [ V \lo T - E ( V \lo T | X, T ) ] \right) + E [ E ( V \lo T | X, T = t)]. 
\end{align*} 
\end{corollary}

Mimicking the limit form of the efficient influence function given in Corollary \ref{corollary:limit form of phi}, we now propose the following kernel-based doubly-debiased machine learning estimator as
\begin{align}
    \label{est:DR}
\hat \vartheta \lo {t, DR} = \frac{1}{n}\sum_{i=1}^n \left\{  \widehat E ( V \lo T | X, T )  + \frac{K\lo h(T\lo i -t)}{\hat{f}\lo{T|X}(t|X\lo i)}\left[V\lo i-   \widehat E ( V \lo T | X, T ) \right] \right\}. 
\end{align}
Again, note that the observable $V \lo T$, rather than the unobservable $V \lo t$, appears in the above equation. 
 The two sample-level conditional expectation  $\widehat E (V \lo T |X, T )$ are computed  by performing nonparametric regression:  
${ \argmin_{v \in \rho ( \ca  Y) } \ \widehat  E (\| V \lo i - v \| \lo {\ca H} \hi 2  | X \lo i , T \lo i )} $
by smoothing spline, kernel regression, or RKHS. The estimation strategies for these infinite-dimensional nuisance parameters are discussed in more detail in Section 4.

\begin{proposition}
    $\hat \vartheta \lo {t, DR} $ in~\eqref{est:DR} is doubly robust, that is, $\E[\hat \vartheta \lo {t, DR} ] = \rho(\beta\lo t)$, even if either of the nuisance parameters $f_{T|X}$ or $\gamfull$, but not both at the same time, is misspecified. 
\end{proposition}

The above proposition shows that 
\begin{align*} 
\E\left[\tilde{\gamma}(t,X) \right] + \lim \lo {h\to 0} E \left[\frac{K_h(T -t)}{f\lo{T|X}(T|X)}[V\lo T - \tilde{\gamma}(T,X)]\right] \nonumber=\vartheta \lo t,  
\end{align*}
from which it is easily deduced that our proposed estimator 
$\hat \vartheta \lo {t, DR}$ is consistent even if $\gamma (t,x)$ is misspecified as $\tilde \gamma (x,t)$. The proof requires the structures of the latent Hilbert space, the law of iterated expectation, and the conditional independence assumption, and can be found in Supplementary Material.

\subsection{Cross-fitting estimator  for  CTROCIN}
While the doubly robust estimator is useful, it requires strong and unverifiable assumptions on the metric space, such as the Donsker property for the function class of the outcome regression. To circumvent these constraints while preserving the desired asymptotic properties, we adopt a sample splitting strategy proposed in~\cite{cher:18} and~\cite{cola:22}. Building on this idea, let $\hgamfull[t,x]{l}$ and $\hpropfull[t,x]{l}$ be suitable estimators based on $(\hat{V}_i, T_i,X_i)_{i=1}^n$ for $\gamfull$ and $f_{T|X}(t|x)$, respectively, and we propose a kernel-based estimator that utilizes the following double-debiased moment function and a cross-fitting strategy.
\begin{enumerate}
    \item[\underline{Step 1}.] Fix $L\in \{2,3,\dots,n\}$ and randomly partition the observation indices into $L$ distinct groups, $I\lo l$, $l=1,2,\dots,L.$
    \item[\underline{Step 2}.] For each $l,$ the estimators for $\gamfull$ and the propensity score given by $f\lo{T|X}(t|x)$ are calculated using observations not present in $I\lo l$ such that these estimates are asymptotically consistent for their respective targets with a suitable rate of convergence.
     \item[\underline{Step 3}.] The proposed doubly robust estimate is given by
     \begin{align}
     \label{cf:est}
        \hat \vartheta\lo {t; CF} = \frac{1}{L}\sum\lo{l=1}\hi L \sum \lo{i\in I\lo l}\left( \hgamfull[t,X\lo i]{l} + \frac{K\lo h(T\lo i -t)}{\hpropfull[t|X_i]{l}}\left[\hat{V}_i -  \hgamfull[t,X\lo i]{l} \right] \right) .
     \end{align}
\end{enumerate}

\section{Theory}
\label{sec:theory}

In this section, we first describe the estimation of the auxiliary quantities involved in~\eqref{cf:est} and then derive results regarding the  asymptotic distribution for the proposed estimator in~\eqref{cf:est}. Before proceeding, we define the following norms: 
\begin{enumerate}
\item for any $f:\ca X \to \real$,  $\| f \| \lo {2, P \lo X} \hi 2 = \int\lo{\ca X} f \hi 2 (x)  dF_X(x)$; \vspace{-.09in}
\item for any $g:\ca T\to \real$, 
$\| g \| \lo {2, P \lo T} \hi 2  := \int \lo {\ca T}g \hi 2 (t) f \lo T (t) d t$; \vspace{-0.09in}
\item  for any $h:\ca T \times \ca X \to \ca H$, $\| h \| \lo {2, P \lo {XT}} \hi 2   =\int \lo {\ca T} \int\lo{\ca X} \|h(x,t)\| \lo {\ca H} \hi 2 f_{TX}(t,x) dxd t $  where $f \lo {TX}$ is the joint density of $(X,T)$.
\end{enumerate}

First, the outcome curves $\hat{V}_i$ (or $\hat Y \lo i$) need to be constructed from the  discrete observations on $V_i$ (or $Y \lo i$), $i=1,\dots, n.$
About these estimates, we make the following convergence assumptions.

 \begin{assump}
 \label{ass:response_est}\textbf{Consistent recovery of the response curves}: The estimates $\hat{Y}_i$; $i=1,\dots,n$ are independent, and there are two real sequences $\alpha_n = o(1)$ and $\nu_n = o(1)$ such that
 \begin{align*}
    & \sup\lo{i=1,\dots,n}\sup\lo{y\in \ca  Y} \E[d\lo Y\hi2(\hat{Y}_i,Y_i)|Y_i = y]  = \sup\lo{i=1,\dots,n}\sup\lo{v\in \rho(\ca  Y)} \E[\|\hat{V}_i-V_i\|^2\lo{\ca H}|V_i = v] = O(\alpha_n^2),\\
    & \sup\lo{i=1,\dots,n}\sup\lo{y\in \ca  Y} \Var [d\lo Y\hi2(\hat{Y}_i,Y_i)|Y_i = y] = \sup\lo{i=1,\dots,n}\sup\lo{v\in \rho(\ca  Y)} \Var[\|\hat{V}_i-V_i\|\lo{\ca H}|V_i = v] = O(\nu_n^4).
 \end{align*}
 \end{assump}

\begin{proposition}
\label{prop:est:outcome:rate}
Under Assumption~\ref{ass:response_est}, we have
\begin{align*}
   \frac{1}{n} \sum_{i=1}^nd^2(\hat{Y}_i,Y_i) = O_P(\alpha_n^2 + \nu_n^2). 
\end{align*}
Further, if $\hat{g}_\oplus$ denotes the empirical Fr\'echet mean of the estimated random objects $\hat{Y}_1,\dots,\hat{Y}_n$ and $\tilde{g}_\oplus$ denotes the empirical Fr\'echet mean of the (not fully) observed objects $Y_1,\dots,Y_n$, i.e.,
\begin{align*}
    \hat{g}_\oplus = \argmin\lo{y\in \ca  Y}\frac{1}{n}\sum_{i=1}^n d\lo Y^2(\hat{Y}_i,y),\ \tilde{g}_\oplus = \argmin\lo{y\in \ca  Y} \frac{1}{n}\sum_{i=1}^n  d\lo Y^2(Y_i,y),
\end{align*}
then
\begin{align}
\label{fr:est:rate}
    d\lo Y^2(\hat{g}_\oplus,\tilde{g}_\oplus) = O_P(\alpha_n^2),\quad  d\lo Y^2(\tilde{g}_\oplus,g_\oplus) = O_P(n^{-1}), 
\end{align}
where $g_\oplus = \argmin\lo{y\in \ca  Y}\E[d\lo Y^2(Y,y)].$
\end{proposition}
Proposition \ref{prop:est:outcome:rate} ensures consistent estimation of the outcome trajectories from the data that are not fully observed. 
Next, we establish the asymptotic distribution of the proposed estimate $\hat \vartheta \lo t$. To do so,  we make some key assumptions about 
the convergence rates of the estimated GPS and conditional expectation, and so on, and describe some existing estimation procedures of $f \lo {T|X}$ and $\gamma (t,x)$. 

We further require the following assumption:
  \begin{assump}
  \label{ass:aux_est} \textbf{Consistent estimation of propensity score and conditional expectation}: For each $l=1,\dots,L,$ and for any $t\in \ca T,$
    \begin{align*}
        & \| \tilde \gamma \lo l (t, \cdot) - \gamma \lo l (t, \cdot)  \| \lo {2, P \lo X} :=
        \left(\int_{\mathcal{X}} (\tgamfull[t,x]{l} - \gamfull)^2 f_{TX}(t,x) dx\right)^{1/2} = O_P(\rho_\gamma),\\ 
      & \| f \lo l (t, \cdot) - f\lo {T|X} (t, \cdot) \| \lo {2,  P \lo X} := \left(\int_{\mathcal{X}} (\hpropfull[t,x]{l} - f_{T|X}(t,x))^2 f_{TX}(t,x) dx\right)^{1/2} = O_P(\rho_f),
    \end{align*}
    where $\rho \lo \gamma,\rho \lo f \to 0$ as $n \to \infty$..
\end{assump}

We discuss the estimation of two nuisance parameters. For the consistent estimation of the quantities involved in~\eqref{cf:est}, we can use any suitable estimates of the conditional Fr\'echet mean dose-response $m(t,X)$
and the GPS $f_{T|X}(t|x)$. For example, one can employ any local or global object regression method to estimate the above using data in the observations not present in the $l$-th bin (as described in Step 2 of the algorithm), $l\in\{1,\dots, L\}$. We discuss some available options in the literature as follows:
\begin{itemize}
    \item The Fr\'echet regression~\citep{pete:19} extends linear least squares and local linear regression to estimate the conditional Fr\'echet mean. As the dimension of the predictor gets higher, its accuracy drops significantly--a phenomenon known as the curse of dimensionality. Later,~\cite{zhan:21} studied the dimension reduction for Fr\'echet regression, and~\cite{bhat:22} introduced single-index Fr\'echet regression by projecting multivariate predictors onto a specified direction 
     to create a single index, thus facilitating more parsimonious Fr\'echet regression.
    \item  The random forest-weighted local Fr\'chet regression was introduced by~\cite{qiu:22}, by employing locally adaptive kernels from random forests as weights to estimate the conditional Fr\'echet mean via locally constant and locally linear averages. Recently, a nonlinear global Fr\'echet regression framework was proposed by~\cite{bhat:23} to extend nonlinear regression to metric spaces and accommodate various model complexities, including linear, polynomial, and function classes dense in $L^2$.
\end{itemize}

To simplify our understanding in the transformed random variables now taking values in a Hilbert space, we can also view the transformed conditional mean $\gamfull = \rho(m(t,X))$ as a unique Riesz representation. As the least squares regression fits nicely into the Hilbert space setting, the conditional mean can be perceived as the best linear predictor using the orthonormal basis of the space once the transformation is done~\citep{rosi:01}.

The estimation of the GPS $f_{T|X}(t|x)$, on the other hand, is a well-studied problem: available estimating procedures include a re-weighted Nadaraya-Watson or locally linear estimator~\citep{fan:96}, 
orthogonal series estimators~\citep{whit:58,wats:69}, penalized quantile regression methods~\citep{bell:19, cola:22}, 
neural networks~\citep{mcca:13,roth:19}, and so on. Essentially, any estimators that satisfy Assumption~\ref{ass:aux_est} would be a candidate for the nuisance parameters estimation from the underlying outcome curves $V_i$, $i=1,\dots,n$.
To tie everything together, the conditional Fr\'echet mean estimator of $\gamfull$ based on the unobserved underlying quantities $V_i$'s and that based on the estimated trajectories $\hat{V}_i$'s are required to be asymptotically close:
\begin{assump}
\label{ass:aux_est2} 
$\| \hat \gamma \lo l ( t, \cdot) - \tilde \gamma (t, \cdot) 
\| \lo {2, P \lo X} := 
\left(\int_{\mathcal{X}} (\hgamfull[t,x]{l} - \tgamfull{l})^2 f_{TX}(t,x) dx\right)^{1/2} = O_P(\alpha^2_n + \nu_n^2) \overset{P}{\longrightarrow}0.
$ 
\end{assump}

Furthermore, the following technical condition is required for deriving the asymptotically linear representation and asymptotic normality for the proposed DML cross-fitting estimator. 

\begin{assump}
\label{ass:differentiability} \textbf{Smooth functions of the treatment effect}: $\E(V |T = t, X = x)$ and $\Var(V |T = t, X = x)$ are three-times differentiable w.r.t. $t$ with all
three derivatives being bounded uniformly over $(v, t, x) \in \mathcal{Z},$ that is there are positive constants $a_k, b_k,\ k=1,2,3$, such that $\sup_{x \in \ca X} \frac{d\hi k}{dt\hi k} |\E(V |T = t, X = x)| \leq a_k$.  and $\sup_{x \in \ca X} \frac{d\hi k}{dt\hi k} \Var(V |T = t, X = x) \leq b_k$. 
\end{assump}

\begin{theorem}
    Under Assumptions~\ref{ass:unconfoundedness}--\ref{ass:aux_est2}, for any $t\in \ca T$,
    \begin{align*}
        \sqrt{nh}(\hat \vartheta\lo {t; CF} - \rho(\beta\lo t)) = \sqrt{nh}\left[\frac{1}{n}\sum_{i=1}^n \frac{K_h(T_i -t)}{f\lo{T|X}(t|X_i)}[V_i - \gamfull[t,X_i]]  + \gamfull[t,X_i] -\rho(\beta\lo t) \right] +o_P(1).
    \end{align*}
Further, if $\E[\lVert V-\gamfull[t,X]\rVert\lo{2,P\lo{XV}}\hi 3|T=t, X]$ and its derivatives w.r.t. $t$ to be bounded uniformly over $(t,x) \in \ca T\times \ca X$. Let $\int\lo{-\infty}\hi {\infty} k\hi 3(u) du <\infty.$ Then by Lyapounov CLT,
\begin{align*}
    \sqrt{nh} (\hat \vartheta\lo {t; CF} -\rho(\beta\lo t) -h\hi 2B\lo t) \overset{\ca L}{\rightarrow} N(0,\Sigma\lo t),
\end{align*}
where $B\lo t := \left(\int\lo{-\infty}\hi{\infty}u\hi2 k(u) du\right)\left[\E\left(\frac{d\hi 2\gamfull[t,X]}{dt\hi 2}\right) + \E\left(\frac{d \gamfull[t,X]}{dt} \frac{ d f\lo{T|X}(t|X)}{dt}/f \lo{T|X}(t|X) \right)\right]$ and\\
$\Sigma\lo t = (const.) \E\left[\Var(V|T=t,X)/f_{T|X}(t|X)\right]$.
\end{theorem}

The above theorem yields an asymptotic inferential framework for the FERF 
and can also be used to quantify the average causal effect between two treatment levels. We define the causal effect map between treatment levels $t$ and $t'$ as $\Delta\lo{tt'} = \rho(\beta\lo t) - \rho(\beta\lo{t'})$.  An exact finite sample guarantee for the uncertainty quantification of the estimators $\hat{\Delta}\lo{tt'}$ can be obtained using the adaptive HulC method by~\cite{kuch:23} to construct confidence regions for the contrast $\Delta\lo{tt'}$, with the following implementation.

Let $\{S_b\}_{b=1}^{B}$ be a (random) partition of $\{1, \dots, n\}$ into $B$ subsets, and the estimators $\hat{\Delta}\lo{tt'}\hi{b} := \{\hat \vartheta\lo {b;t}) - \hat \vartheta\lo{b;t'})\}_{b=1}^{B}$ for $ \Delta\lo{tt'} =\rho(\beta\lo t) - \rho(\beta\lo{t'})$ computed for each subsample $\{(V_i, T_i, X_i) : i \in S_b\}_{b=1}^{B}$. Define the maximum median bias of the estimators $\{\hat{\Delta}\lo{tt'}\hi{b}\}_{b=1}^{B}$  for $\Delta\lo{tt'}$ as

\[
\Delta := \max_{1 \leq b \leq B} \left\{ 0, \frac{1}{2} - \min \left\{ P(U_b \geq 0), P(U_b \leq 0) \right\} \right\},
\]
where $U_b = \hat{\Delta}\lo{tt'}\hi{b} - \Delta\lo{tt'}$.
Adopting the HulC algorithm, we construct a confidence interval with coverage probability $1 - \alpha$ for $\Delta\lo{tt'}$ as follows:

\begin{enumerate}
    \item Find the smallest integer $B = B_{\alpha, \Delta} \geq 1$ such that
    \[
    P(B; \Delta) := \left( \frac{1}{2} - \Delta \right)^B + \left( \frac{1}{2} + \Delta \right)^B \leq \alpha.
    \]
    
    \item Generate a uniform random variable $U \in [0, 1]$ and set $B^*$ as:
    \[
    B^* := \begin{cases} 
    B_{\alpha, \Delta} & \text{if } U \leq \tau_{\alpha, \Delta}:= \frac{\alpha - P(B_{\alpha, \Delta}; \Delta)}{P(B_{\alpha, \Delta} - 1; \Delta) - P(B_{\alpha, \Delta}; \Delta)}, \\
    B_{\alpha, \Delta} - 1 & \text{otherwise}.
    \end{cases}
    \]
    
    \item Randomly split $\{(V_i, T_i, X_i)\}_{i=1}^n$ into $B^*$ disjoint sets $\{(V_i, T_i, X_i) : i \in S_b\}_{b=1}^{B^*}$ and compute $\hat{\Delta}\lo{tt'}\hi{b} := \{\hat \vartheta\lo {b;t}) - \hat \vartheta\lo{b;t'})\}_{b=1}^{B^*}$.
    
    \item Compute the confidence interval:
    \[
    \hat{C}_{\alpha, \Delta} := \left[ \min_{1 \leq b \leq B^*} \hat{\Delta}\lo{tt'}\hi{b}, \max_{1 \leq b \leq B^*} \hat{\Delta}\lo{tt'}\hi{b}  \right].
    \]
\end{enumerate}
From Theorem 1 in~\cite{kuch:23}, the coverage probability of this confidence interval is guaranteed for finite samples, i.e.,
$
P\left( \Delta\lo{tt'} \in \hat{C}_{\alpha, \Delta} \right) \geq 1 - \alpha.$

\section{Simulation studies}
\label{sec:simu}
The four estimators -- Outcome Regression (OR), Inverse Probability Weighting (IPW), Doubly-Robust (DR), and Doubly-Debiased Cross Fitting (CF) -- are introduced in logical progression in Section 3. To assess their performance, we conduct simulations across various settings involving different types of metric space-valued responses: univariate distributions equipped with the Wasserstein metric, covariance matrices equipped with
the Frobenius metric, and six-dimensional compositional data on the positive segment of the sphere $\ca S\hi 6 \lo +$ equipped with the geodesic metric on $\ca S\hi 6$. It is important to note that we do not have any existing methods to compare with since the CTROCIN framework is mostly unexplored as of now. We consider sample sizes $n= 50,200,$ and $1000$ and report the average and standard deviation of leave-one-out mean squared error (MSE) over $B= 100$ Monte Carlo replications. For the $b^{\text{th}}$ simulation, the MSE is given by
\[
\text{MSE}\hi{(b)} = \frac{1}{n} \sum_{i=1}^n d^2(Y_i, \hat{Y}_{(i)}),
\]
where $Y_i$ is the observed response for the $i^{\text{th}}$ sample, $\hat{Y}_{(i)}$ is the predicted value for $Y_i$ obtained by fitting the model on the remaining $n-1$ observations and predicting $Y_i$. To compute the MSE efficiently, each response $Y_i$ is first embedded into a Hilbert space via a known isometric map $\rho$, resulting in $V_i = \rho(Y_i)$. This transformation allows MSE to be computed as squared distances using the inner product in the embedding space.

In all simulations, the confounder or pre-exposure covariates $X$ and the treatment variable conditioned on $X$ are generated as follows:
We generate six pre-exposure covariates $(X_1, X_2,\dots, X_6)$ as a combination of continuous and categorical variables:
$$X_1,\dots,X_4 \sim N(0,I_4),\ X_5 \sim V\{-2,2\},\ X_6 \sim U(-3,3),$$
where $N(0,I_4)$ denotes a multivariate normal distribution, $V\{-2,2\}$ denotes a discrete uniform distribution, and $U(-3,3)$ denotes a uniform distribution. We generate $T$ using three different specifications of the GPS model, all relying on the cardinal function $r(X) = - 0.8 + (0.1,0.1,-0.1,0.2,0.1,0.1)^\top X$. The coefficients of the cardinal function $r(X)$ are modified from~\cite{wu2024matching}. Specifically, we consider the following scenarios: 
\begin{enumerate}
    \item $T = 0.9r(X) +1+N(0,.5)$
    \item $T = 0.5 r(X) + 0.2 + T(2)$
   \item $ T = 0.7 \log(r(X)) + 1.3 + N(0,1)$
\end{enumerate}
Scenario 1 serves as the baseline,  where the exposure $T$ is generated as a linear function of the confounders, and the residuals are normally distributed without extreme values. Scenario 2 introduces heavy-tailed behavior by generating $T$ from a t-distribution, leading to extreme values and, consequently, extreme GPS values. Scenario 3 can be seen as a variant of scenario 1 by incorporating a more complex data-generating process and deliberately misspecifying the GPS model, thereby testing robustness under model misspecification.


For space considerations, we only present simulation results for distributional responses. The additional simulation results for SPD matrix objects and compositional data taking values on the surface of the sphere $S^2 \subset \real^3$ can be found in the supplement.

We generate $Y$ from an outcome model that is assumed to depend on the treatment and confounders. To this end, we consider two settings as follows. 
\begin{enumerate}
    \item[(A)] $Y$ is  a distributional object whose mean is  cubic function of $T$ with additive terms for confounders and interactions between $T$ and confounders,
$Y |T,X \equiv N(\gamma(T,X), \sigma^2)$, where $\gamma(T,X)= 1 -(0.2, 0.2,0.3,-0.1,0.2,0.2)^\top X - T(0.1 - 0.1X_1 + 0.1X_4 + 0.1X_5 +0.1X_3^2)+0.1T^3$, and $\sigma^2=1$.
\item[(B)] The random transport maps $T$ 
are generated by sampling uniformly from the collection of transport maps $T_k(x) =  x - \sin(k\pi x)/|k \pi|$, for $k \in \{-2,-1,1,2\}$, with
$Y = T \circ(\gamma(T,X) + \sigma \Phi^{-1})$, where $\Phi(\cdot)$ is the standard normal distribution function. 
\end{enumerate} 
Setting (A) assumes Gaussian outcomes, while Setting (B) introduces greater complexity by generating non-Gaussian distributions. In Setting (B), distribution parameters are first sampled as in Setting (A), but the resulting distributions are then ``transported'' in Wasserstein space following~\cite{pete:19} and \cite{chen:22}.

The simulated outcomes, represented as quantile functions, are embedded in the Hilbert space $L\hi2 [0,1]$ and taken as the effective outcomes. The estimated  GPS $\hat{f}_{t|X}$ is computed by a cross-validated Super Learner ensemble algorithm, implemented by the R package \texttt{SuperLearner} with the extreme gradient boosting machines algorithm SL.xgboost. The other outcome regression function $\gamfull$ is estimated using smoothing splines via the R function \texttt{smooth.spline}, employing its default generalized cross-validation for tuning.

\begin{table}[!htb]
\centering
\resizebox{\columnwidth}{!}{%
\begin{tabular}{cc||ccc||ccc||ccc||ccc}
\hline
 &  & \multicolumn{3}{c||}{OR} & \multicolumn{3}{c||}{IPW} & \multicolumn{3}{c||}{DR} & \multicolumn{3}{c}{CF} \\ \hline\hline
\multicolumn{1}{c||}{\begin{tabular}[c]{@{}c@{}}GPS \\ Scenario  \end{tabular}} & \begin{tabular}[c]{@{}c@{}}Outcome\\ Model\end{tabular} & \multicolumn{1}{c|}{$n = 50$} & \multicolumn{1}{c|}{$n = 200$} & $n = 1000$ & \multicolumn{1}{c|}{$n= 50$} & \multicolumn{1}{c|}{$n  = 200$} & $n = 1000$ & \multicolumn{1}{c|}{$n = 50$} & \multicolumn{1}{c|}{$n = 200$} & $n = 1000$ & \multicolumn{1}{c|}{$n = 50$} & \multicolumn{1}{c|}{$n =200$} & $n = 1000$ \\ \hline
\multicolumn{1}{c||}{\multirow{2}{*}{1}} & A & \multicolumn{1}{c|}{\begin{tabular}[c]{@{}c@{}}2.39\\ (1.67)\end{tabular}} & \multicolumn{1}{c|}{\begin{tabular}[c]{@{}c@{}}1.33\\ (0.88)\end{tabular}} & \begin{tabular}[c]{@{}c@{}}0.34\\ (0.11)\end{tabular} & \multicolumn{1}{c|}{\begin{tabular}[c]{@{}c@{}}3.79\\ (2.11)\end{tabular}} & \multicolumn{1}{c|}{\begin{tabular}[c]{@{}c@{}}1.12\\ (2.09)\end{tabular}} & \begin{tabular}[c]{@{}c@{}}0.57\\ (1.81)\end{tabular} & \multicolumn{1}{c|}{\begin{tabular}[c]{@{}c@{}}1.67\\ (0.92)\end{tabular}} & \multicolumn{1}{c|}{\begin{tabular}[c]{@{}c@{}}0.58\\ (0.78)\end{tabular}} & \begin{tabular}[c]{@{}c@{}}0.09\\ (0.11)\end{tabular} & \multicolumn{1}{c|}{\begin{tabular}[c]{@{}c@{}}1.67\\ (0.41)\end{tabular}} & \multicolumn{1}{c|}{\begin{tabular}[c]{@{}c@{}}0.60\\ (0.39)\end{tabular}} & \begin{tabular}[c]{@{}c@{}}0.10\\ (0.21)\end{tabular} \\ \cline{2-14} 
\multicolumn{1}{c||}{} & B & \multicolumn{1}{c|}{\begin{tabular}[c]{@{}c@{}}3.12\\ (1.33)\end{tabular}} & \multicolumn{1}{c|}{\begin{tabular}[c]{@{}c@{}}1.34\\ (1.02)\end{tabular}} & \begin{tabular}[c]{@{}c@{}}0.39\\ (0.81)\end{tabular} & \multicolumn{1}{c|}{\begin{tabular}[c]{@{}c@{}}4.01\\ (2.11)\end{tabular}} & \multicolumn{1}{c|}{\begin{tabular}[c]{@{}c@{}}2.89\\ (1.89)\end{tabular}} & \begin{tabular}[c]{@{}c@{}}1.13\\ (1.33)\end{tabular} & \multicolumn{1}{c|}{\begin{tabular}[c]{@{}c@{}}1.45\\ (1.08)\end{tabular}} & \multicolumn{1}{c|}{\begin{tabular}[c]{@{}c@{}}0.53\\ (1.03)\end{tabular}} & \begin{tabular}[c]{@{}c@{}}0.17\\ (0.71)\end{tabular} & \multicolumn{1}{c|}{\begin{tabular}[c]{@{}c@{}}1.45\\ (1.03)\end{tabular}} & \multicolumn{1}{c|}{\begin{tabular}[c]{@{}c@{}}0.46\\ (0.71)\end{tabular}} & \begin{tabular}[c]{@{}c@{}}0.09\\ (0.11)\end{tabular} \\ \hline
\multicolumn{1}{c||}{\multirow{2}{*}{2}} & A & \multicolumn{1}{c|}{\begin{tabular}[c]{@{}c@{}}3.33\\ (1.14)\end{tabular}} & \multicolumn{1}{c|}{\begin{tabular}[c]{@{}c@{}}2.78\\ (1.11)\end{tabular}} & \begin{tabular}[c]{@{}c@{}}1.65\\ (0.51)\end{tabular} & \multicolumn{1}{c|}{\begin{tabular}[c]{@{}c@{}}5.39\\ (2.11)\end{tabular}} & \multicolumn{1}{c|}{\begin{tabular}[c]{@{}c@{}}3.23\\ (2.10)\end{tabular}} & \begin{tabular}[c]{@{}c@{}}2.72\\ (2.05)\end{tabular} & \multicolumn{1}{c|}{\begin{tabular}[c]{@{}c@{}}2.33\\ (1.66)\end{tabular}} & \multicolumn{1}{c|}{\begin{tabular}[c]{@{}c@{}}1.36\\ (1.67)\end{tabular}} & \begin{tabular}[c]{@{}c@{}}0.91\\ (0.88)\end{tabular} & \multicolumn{1}{c|}{\begin{tabular}[c]{@{}c@{}}2.25\\ (1.03)\end{tabular}} & \multicolumn{1}{c|}{\begin{tabular}[c]{@{}c@{}}1.35\\ (0.91)\end{tabular}} & \begin{tabular}[c]{@{}c@{}}0.89\\ (0.41)\end{tabular} \\ \cline{2-14} 
\multicolumn{1}{c||}{} & B & \multicolumn{1}{c|}{\begin{tabular}[c]{@{}c@{}}4.02\\ (1.60)\end{tabular}} & \multicolumn{1}{c|}{\begin{tabular}[c]{@{}c@{}}2.88\\ (0.98)\end{tabular}} & \begin{tabular}[c]{@{}c@{}}1.77\\ (1.21)\end{tabular} & \multicolumn{1}{c|}{\begin{tabular}[c]{@{}c@{}}9.21\\ (2.33)\end{tabular}} & \multicolumn{1}{c|}{\begin{tabular}[c]{@{}c@{}}5.34\\ (2.08)\end{tabular}} & \begin{tabular}[c]{@{}c@{}}3.34\\ (1.56)\end{tabular} & \multicolumn{1}{c|}{\begin{tabular}[c]{@{}c@{}}2.89\\ (1.03)\end{tabular}} & \multicolumn{1}{c|}{\begin{tabular}[c]{@{}c@{}}1.90\\ (1.02)\end{tabular}} & \begin{tabular}[c]{@{}c@{}}1.22\\ (0.99)\end{tabular} & \multicolumn{1}{c|}{\begin{tabular}[c]{@{}c@{}}2.30\\ (1.31)\end{tabular}} & \multicolumn{1}{c|}{\begin{tabular}[c]{@{}c@{}}1.81\\ (1.01)\end{tabular}} & \begin{tabular}[c]{@{}c@{}}0.79\\ (0.41)\end{tabular} \\ \hline
\multicolumn{1}{c||}{\multirow{2}{*}{3}} & A & \multicolumn{1}{c|}{\begin{tabular}[c]{@{}c@{}}3.56\\ (1.45)\end{tabular}} & \multicolumn{1}{c|}{\begin{tabular}[c]{@{}c@{}}1.78\\ (1.21)\end{tabular}} & \begin{tabular}[c]{@{}c@{}}0.45\\ (0.11)\end{tabular} & \multicolumn{1}{c|}{\begin{tabular}[c]{@{}c@{}}4.21\\ (1.72)\end{tabular}} & \multicolumn{1}{c|}{\begin{tabular}[c]{@{}c@{}}2.17\\ (1.32)\end{tabular}} & \begin{tabular}[c]{@{}c@{}}0.52\\ (1.12)\end{tabular} & \multicolumn{1}{c|}{\begin{tabular}[c]{@{}c@{}}2.08\\ (0.67)\end{tabular}} & \multicolumn{1}{c|}{\begin{tabular}[c]{@{}c@{}}0.78\\ (0.37)\end{tabular}} & \begin{tabular}[c]{@{}c@{}}0.12\\ (0.36)\end{tabular} & \multicolumn{1}{c|}{\begin{tabular}[c]{@{}c@{}}1.89\\ (0.66)\end{tabular}} & \multicolumn{1}{c|}{\begin{tabular}[c]{@{}c@{}}0.67\\ (0.37)\end{tabular}} & \begin{tabular}[c]{@{}c@{}}0.11\\ (0.37)\end{tabular} \\ \cline{2-14} 
\multicolumn{1}{c||}{} & B & \multicolumn{1}{c|}{\begin{tabular}[c]{@{}c@{}}3.32\\ (1.33)\end{tabular}} & \multicolumn{1}{c|}{\begin{tabular}[c]{@{}c@{}}2.05\\ (1.32)\end{tabular}} & \begin{tabular}[c]{@{}c@{}}0.55\\ (0.79)\end{tabular} & \multicolumn{1}{c|}{\begin{tabular}[c]{@{}c@{}}6.33\\ (2.12)\end{tabular}} & \multicolumn{1}{c|}{\begin{tabular}[c]{@{}c@{}}2.89\\ (1.89)\end{tabular}} & \begin{tabular}[c]{@{}c@{}}1.07\\ (0.98)\end{tabular} & \multicolumn{1}{c|}{\begin{tabular}[c]{@{}c@{}}1.87\\ (0.97)\end{tabular}} & \multicolumn{1}{c|}{\begin{tabular}[c]{@{}c@{}}0.80\\ (0.58)\end{tabular}} & \begin{tabular}[c]{@{}c@{}}0.12\\ (0.11)\end{tabular} & \multicolumn{1}{c|}{\begin{tabular}[c]{@{}c@{}}1.55\\ (0.91)\end{tabular}} & \multicolumn{1}{c|}{\begin{tabular}[c]{@{}c@{}}0.45\\ (0.21)\end{tabular}} & \begin{tabular}[c]{@{}c@{}}0.09\\ (0.12)\end{tabular} \\ \hline
\end{tabular}%
}
\caption{The average  (standard deviation) of the Mean Integrated Square Error (MISE) over 100 Monte Carlo iterations for four estimators: Outcome Regression (OR), Inverse Probability Weighting (IPW), Doubly Robust  (DR), and Cross Fitting (CF). The data-generating mechanisms include three scenarios for generating the GPS $f_{t|X}$, each paired with one of  three outcome models: (A), (B), and (C).}
\label{tab:mise:distribution}
\end{table}

Table~\ref{tab:mise:distribution} compares the performance of the OR, IPW, DR, and CF estimators. Overall, all estimators successfully recover the true outcome distribution. The MSE of the doubly robust estimator decreases as the sample size increases. For more complicated data-generating mechanisms (e.g., combinations of scenario 2 for the propensity score model and outcome generation model B), the MSEs are generally higher. Furthermore, neither the OR nor IPW estimator demonstrates double robustness under model misspecification, displaying large MSEs with high variances even with a sample size of $n =1000$. The DR and CF estimators offer improved estimations. These results are in line with our theoretical analysis.

\begin{table}[!htb]
\centering
{\small \begin{tabular}{c||c||c|c|c}
\hline 
{\small GPS Scenario}& {\small Outcome Model} & {\small $n = 50$} & {\small $n =200$} & {\small $n = 1000$} \\ \hline \hline
\multirow{2}{*}{1} & {\small A} & 2.72 (1.12) &  2.05 ( 1.05) & 1.89 (0.88) \\ \cline{2-5} 
 & {\small B} & 3.01 (1.06) & 2.77 (1.12) & 1.55 (0.67) \\ \cline{2-5} \hline
\multirow{2}{*}{2} & {\small A} & 3.11 (1.12)  & 2.44 (1.13) & 1.77 (0.67) \\ \cline{2-5} 
 & {\small B} & 3.12 (1.11) & 2.79 (0.98) & 1.89 (0.67) \\ \cline{2-5} \hline
\multirow{2}{*}{3} & {\small A} & 2.05 (1.02) & 1.79 (1.01) & 1.56 (0.91) \\ \cline{2-5} 
 & {\small B} &  2.03 (0.99) & 1.89 (0.76) & 1.78 (0.76) \\ \cline{2-5} \hline
\end{tabular}}
\caption{The average (standard deviation) of the Wasserstein distance between the lower and upper confidence bands computed using the proposed Cross-Fitting (CF) estimator over 100 Monte Carlo iterations across different combinations of data-generating mechanisms.}
\label{tab:wass:CI:distribution}
\end{table}

Table~\ref{tab:wass:CI:distribution} shows the average width of the $95\%$ confidence band obtained from the asymptotic properties of the CF estimate under various simulation scenarios and varying sample sizes, again over $100$ Monte Carlo simulation runs. The width of the confidence band generally narrows with increasing sample size.

\section{Air pollution and health application}
\label{sec:data}

The key scientific question in air pollution epidemiology studies is to assess whether and to what extent exposure to air pollution is causally linked to adverse health outcomes. Specifically, we aim to apply our proposed doubly debiased approach to estimate the causal FERF of long-term \pmfine\  \  exposure on all-cause mortality. The response is the age-at-death densities, which reside in the metric space of univariate distributions. Equipped with the Wasserstein-2 metric, this space of distributions can be isometrically embedded into the Hilbert space $L^2[0,1]$. The mortality data for various counties in the United States are obtained from the Centers for Disease Control and Prevention (CDC) website. We use the data for the year $2012$, which is available in the form of cohort life tables over the age interval $[0,110]$ for each of $n = 2392$ counties in the U.S.. For each county, the life table data that correspond to histograms are smoothed with a bin width of five years by adding a smoothing step using the R package fr\'echet~\citep{fr:package} with a bandwidth of 2 years. Thus, a sample of $n =2392$ age-at-death densities is obtained, which are then embedded in the underlying Hilbert space $L^2[0,1]$ to produce the outcomes of interest. 

The treatment data is collected as the county-level long-term exposure to  \pmfine\  \  (averaged from $2000$ to $2016$) from an established exposure prediction model. The resources used are discussed below.
The estimated daily concentrations of \pmfine\  \ are available for 1-km square grids for the contiguous USA between $2000$ and $2016$~\citep{di2019ensemble}. These measurements are generated using an ensemble-based model that fuses predictions from three machine learning methods: a random forest regression, a gradient
boosting machine, and a neural network. Each model uses more than 100 predictor variables derived from satellite data, land-use data, weather measurements, and output from chemical transport
model (CTM) simulations. The ensemble was trained on daily \pmfine\  \  concentrations measured at $2,156$ U.S. EPA monitoring sites, with an average cross-validated $R^2$ of $0.86$ for daily \pmfine\  \  predictions and $0.89$ for annual \pmfine\  \  predictions, indicating excellent performance. These predictions have
been used in previous high-impact studies evaluating \pmfine\  \  and mortality~\citep{jose:23, wei2021emulating}. In short, CTM and satellite data are combined to estimate a high-resolution \pmfine\  \  surface across the whole United States. This surface is bias-corrected for ground-monitor \pmfine\  \  observations using a geographically weighted regression. We aggregated these levels spatially by averaging the values for all grid points within a county to obtain the temporally averaged \pmfine\  \  values (2000–2016) at the county level by averaging estimated \pmfine\  \  values within a given county.

We first estimate the GPS by using an Extreme Gradient Boosting Machine (GBM) (i.e., a single learner in the Super Learner algorithm)~\citep{chen2016xgboost, zhu2015boosting}, with the county-level \pmfine\  \  exposure as the dependent variable and 19 zip code-level potential confounders as independent variables. The latter include population demographics (\%Female, \%Black, \%Hispanic populations, Population density), health-related (Mean BMI, \% Ever Smoked), and educational (\% Below High School Education) information regarding socioeconomic status (Median Home Value, Median Household Income, \% Owner-occupied Housing), and meteorological information (Summer and Winter minimum and maximum temperature, humidity). The extreme GBM learner is desirable for the estimation of the infinite-dimensional nuisance parameter $f_{T|X}$ for 19-dimensional confounder $X$, as described, since the method is flexible, computationally feasible on a large dataset, and achieves better covariate balance compared to a linear regression model on the complex application data.

 \begin{figure}[!htb]
		\centering %
		\begin{subfigure}{0.45\textwidth}
		\centering %
			\includegraphics[width=.9\textwidth]{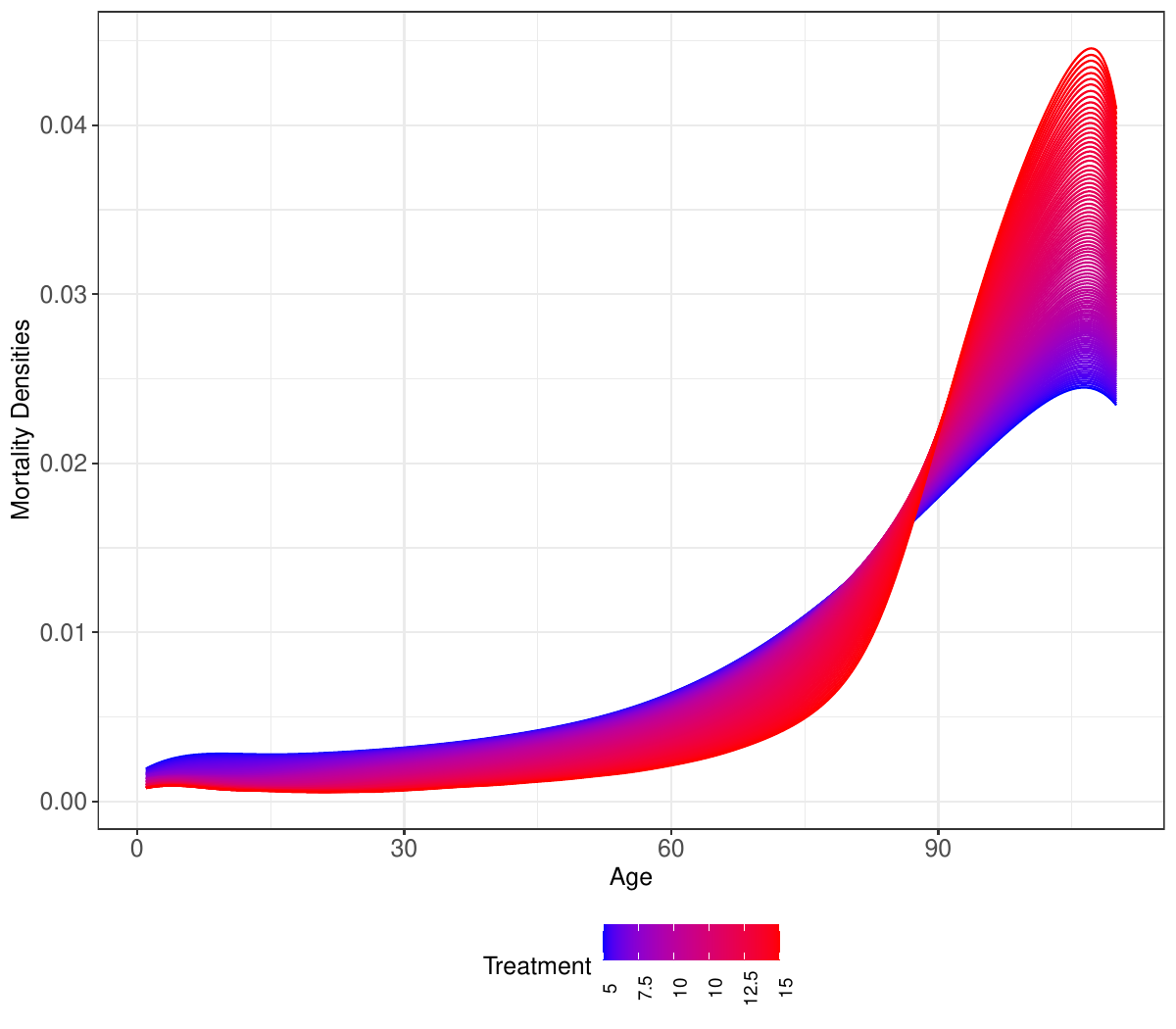}
		\end{subfigure}\hfil 
		\begin{subfigure}{0.45\textwidth}
		\centering %
			\includegraphics[width=.9\textwidth]{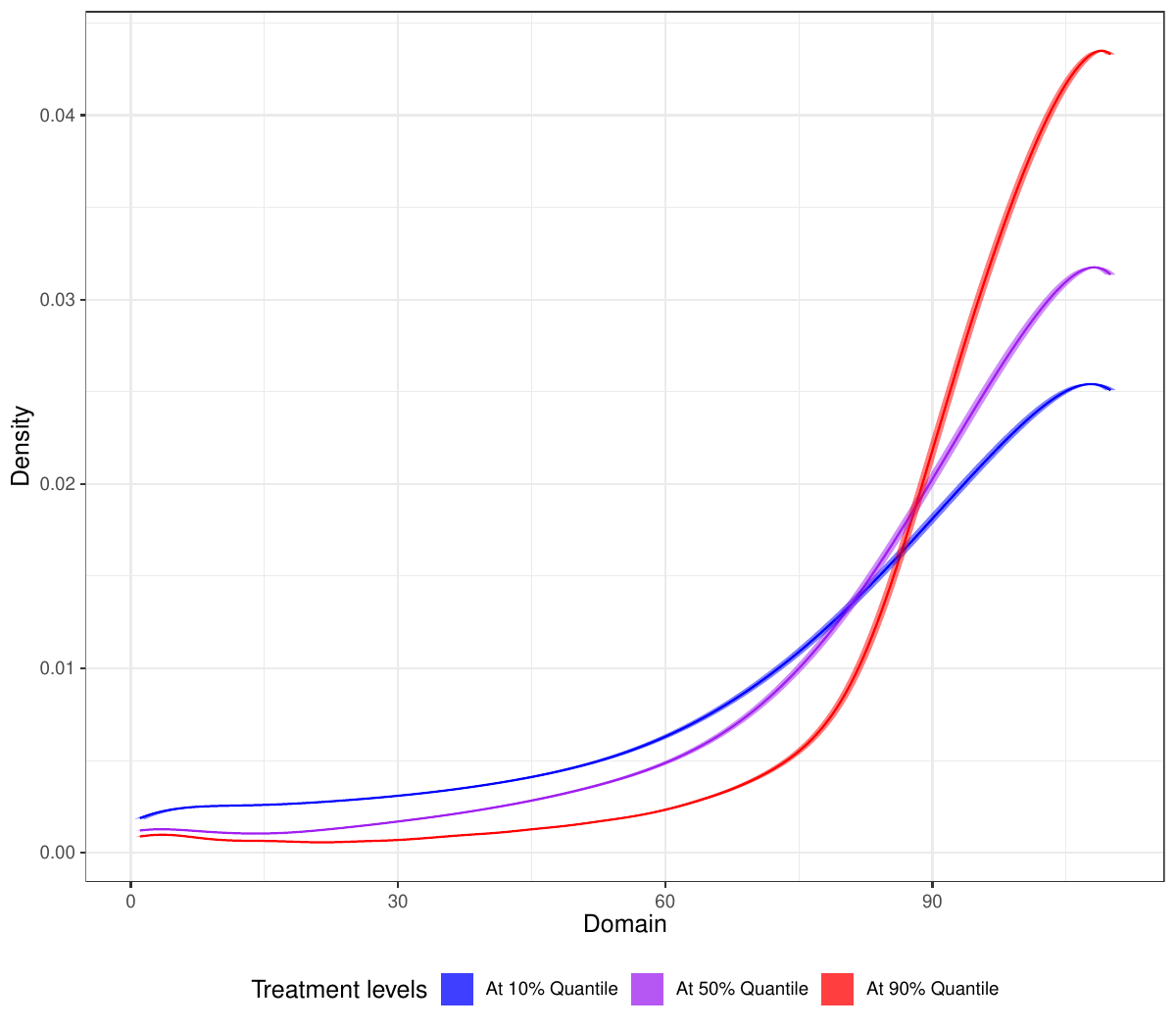}
		\end{subfigure}
		\centering
		\caption{The left panel shows the changes in the density of the age-at-death distribution as the treatment \pmfine \  \ increases from low (blue) to high (red). The right panel presents the 95\% pointwise confidence bands for three different levels of the treatment values, where the red, magenta, and blue lines correspond to the 10\%, 50\%, and 90\% percentiles, respectively.}
		\label{fig:data:mort:densities}
	\end{figure}

Next, for estimating the outcome regression model $E(\rho(Y_t)|T=t, X)$, we implemented a global linear Fr\'echet regression~\citep{pete:19, bhat:23} model using the Hilbertian quantile functions $\rho(Y_t)$ across $n= 2392$ counties as responses and the confounders $X$ as predictors, at every given level $t$ of the treatment, \pmfine\ \ exposure. Finally, we implemented the cross-fitting estimator in~\eqref{cf:est} using $L = 100$. The left panel of Figure~\ref{fig:data:mort:densities} shows the changes in the potential age-at-densities for varying levels of the treatment.  The counterfactual densities are color-coded such that blue to red indicates smaller to larger values of the \pmfine\ \ exposures. 
We find that a lower treatment level can be causally linked with left-shifted age-at-death distributions in general, while higher treatment levels result in a shift of the mode of the age-at-death toward the right. We also plot the $95\%$ pointwise confidence band at three different levels of the \pmfine\ \ exposure, namely, at the $10^{\text{th}}, 50^{\text{th}}$, and $90^{\text{th}}$ percentiles of the treatment values (see the right panel of Figure~\ref{fig:data:mort:densities}). However, the bands are very narrow, which makes the interpretation of the potential outcome distribution rather obscure, possibly indicating a heterogeneous group variation in the treatment effects.

\begin{figure}[!htb]
    \centering
    \includegraphics[width=.75\textwidth]{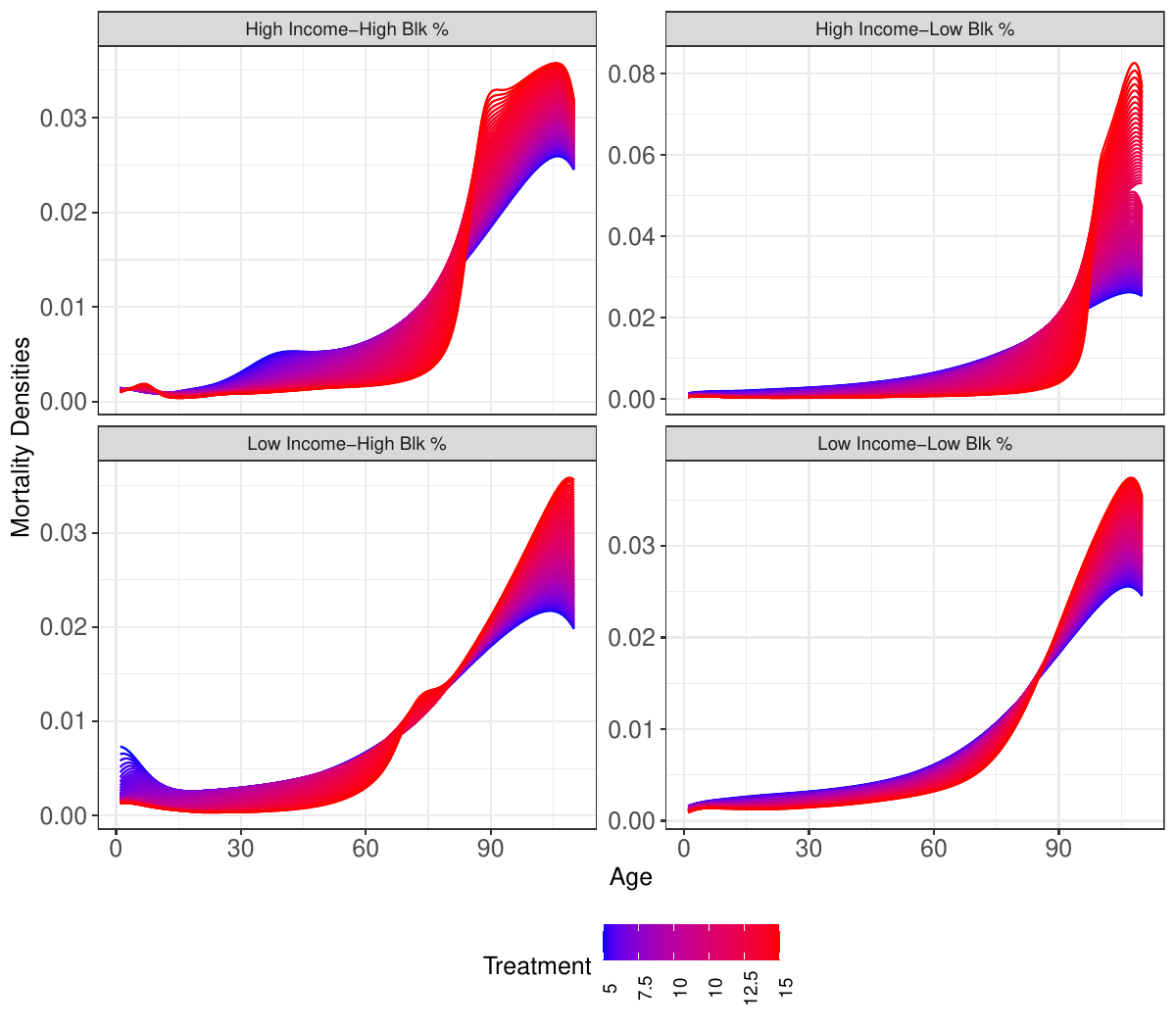}
    \caption{The panels show (clockwise) the changes in the density of the age-at-death distribution as the treatment level, \pmfine\ \ increases from low (blue) to high (red) over the four socio-demographic groups. The low-income-high black \% population appears to be more impacted in terms of higher child mortality and lower age-at-death as the \pmfine\ \ level worsens.}
    \label{fig:data:mort:grp}
\end{figure}


To examine heterogeneity in the causal effect of \pmfine\ \ exposure,  we divide the sample into four disjoint groups based on socio-economic characteristics, following~\cite{jose:23}. Specifically, we stratify counties by combinations of high or low income and high or low percentage of Black residents. The CF estimator is implemented for four groups separately. The resulting counterfactual age-at-death distributions exhibit different shapes and structures across these four groups and resonate with the finding of~\cite{jose:23}. As shown in Figure~\ref{fig:data:mort:densities}, lower \pmfine\ \ exposure is causally associated with lower mortality in the full population, but marginalized subpopulations appear to benefit more as the \pmfine\ \ levels decrease (see Figure~\ref{fig:data:mort:grp}). For example, the mode of the distribution shifts toward the top-right when the \pmfine\ \ levels are lower for the low-income, high-black\% group, indicating a higher longevity corresponding to lower \pmfine\ \ levels. The child mortality is also lower with a lower \pmfine\ \ level. The effect is not so evident for the high-income-low black group.

\begin{figure}[!htb]
    \centering
    \includegraphics[width=.6\textwidth]{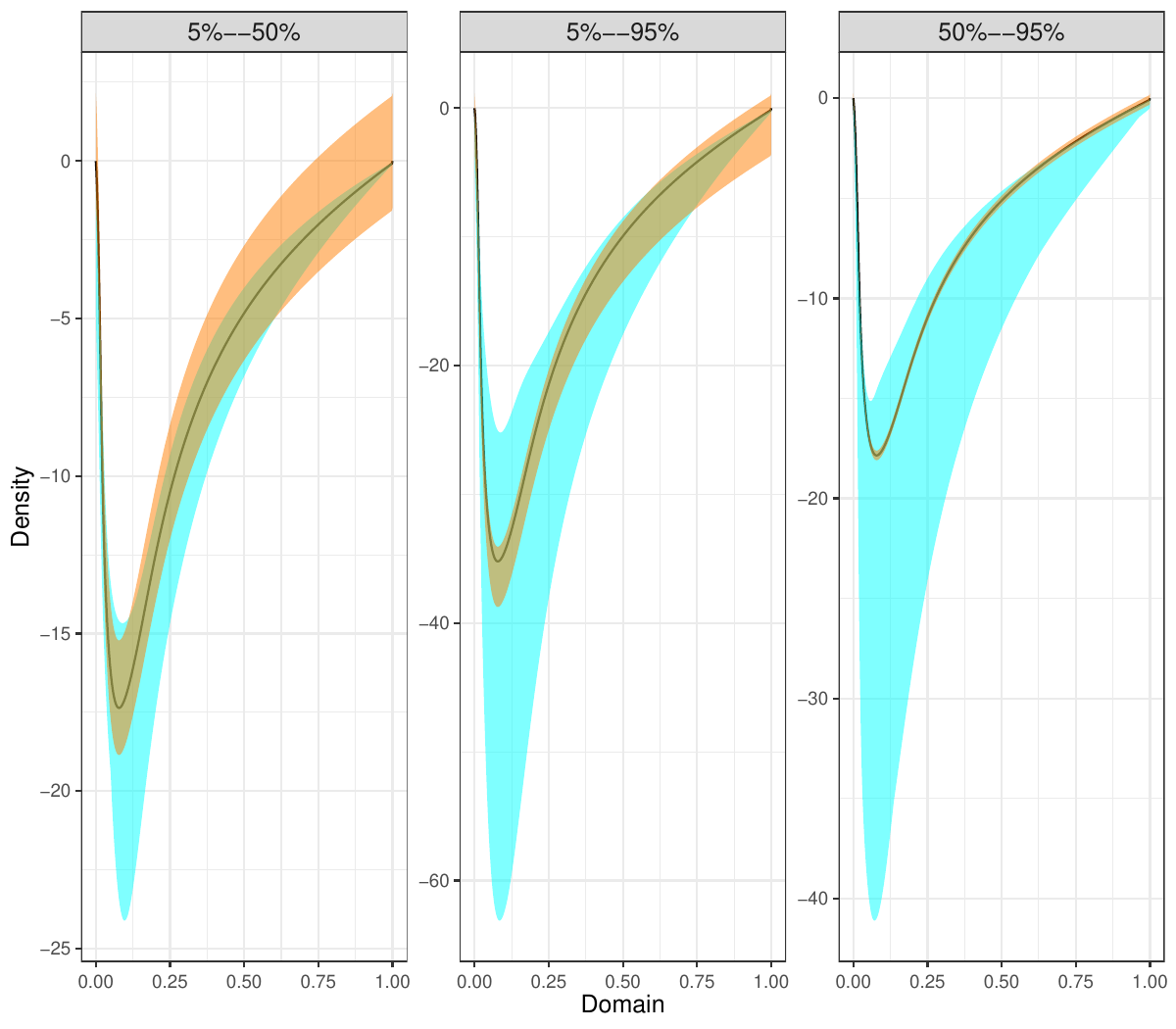}
    \caption{Causal map for pairwise comparison between the treatment levels 5\%, 50\% and 90\% of the \pmfine\  \  values. The dark orange bands are the asymptotic 95\% confidence band derived from our method, and the cyan region is the exact (finite sample) 95\% confidence region derived from implementing the hulC method by~\citep{kuch:23}.}
    \label{fig:data:mort:c_map:total}
\end{figure}

We also compute the causal effect map for pairwise comparisons of three treatment levels (at 5\%, 50\%, and 95\% values of the treatment levels, respectively). We evaluate a $95\%$ pointwise confidence band obtained from the asymptotic distribution of our proposed estimate using Theorem 2. We also implement the $95\%$ pointwise confidence band using the HulC method~\citep{kuch:23}. Figure~\ref{fig:data:mort:c_map:total} shows the causal effect maps for the contrast of $t= 90\%- t' = 50\%$, $t= 50\%- t' = 10\%$, and $t= 90\%- t' = 10\%$, where for each contrast level, the confidence band obtained from the two methods are overlaid. The confidence band derived from the asymptotic distribution (in orange) is tighter than the general method of minimizing maximum bias, as designed in~\cite{kuch:23} (in blue).

\begin{figure}[H]
    \centering
    \includegraphics[width=.75\textwidth]{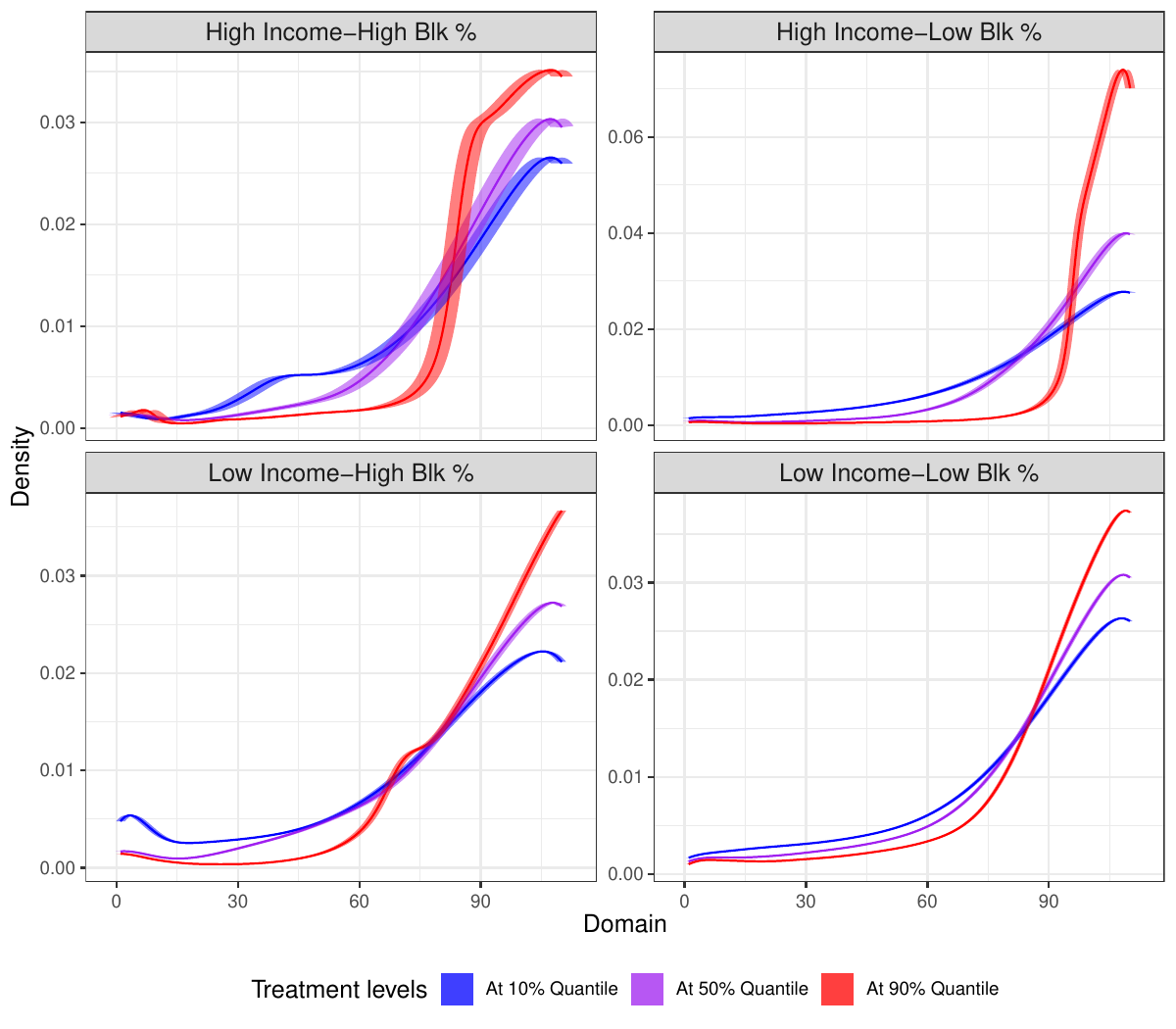}
    \caption{The 95\% pointwise confidence bands for four different socio-demographic groups are displayed for three different levels of the treatment values, where the red, magenta, and blue lines represent the 10\%, 50\%, and 90\% percentiles of the treatment.}
    \label{fig:data:mort:grp:CI}
\end{figure}

Figure~\ref{fig:data:mort:grp:CI} shows the $95\%$ confidence bands for each group. As a conclusion, we might infer that the groups with higher income-higher Black\%, low income-low Black\%, and low income-higher Black\% populations may benefit more from lower \pmfine\ \ levels than higher income low Black\% groups. These findings underscore the importance of considering racial identity and income when assessing health inequities.

\section{Conclusion}
\label{sec:concl}

We proposed a method for estimating the causal effect of a continuous treatment on random object response, assuming that the metric space for the response can be embedded in a Hilbert space. This covers many commonly observed random objects such as distributional data, SPD matrices, data on a Riemannian manifold, etc. However, this embedding assumption imposes certain limitations on the method. For example, in certain cases, it may be impossible to embed the metric space or the form of the embedding map may be unknown. This occurs, for example, in spaces such as phylogenetic trees with the BHV or hyperbolic metric~\citep{bill:01, mata:24}. 
Moreover, the embedding map is not necessarily bijective, making it non-trivial to project back to the original metric space. This highlights the need for intrinsic methods, in which all model components and fits are defined directly within the metric space~\citep{schotz2021frechet, bhat:23}. Alternatively, we can consider a general method for CTROCIN without Hilbert space embedding by focusing on the metric $d  (Y,y)$ rather than the random object $Y$ itself.  This insight provides a useful paradigm for performing Fr\'echet regression in settings where embedding into a Hilbert space is not feasible. In particular, we can define a causally unbiased estimate of the metric $d \hi 2 (Y \lo t, y)$ for each $y \in \ca Y$ using the IPW or doubly robust estimate and allows for a semiparametric efficient (a.k.a. doubly robust) estimate. This calls for a future research agenda in the intersection of random object data analysis and causal inference. 

{\small
\renewcommand{\baselinestretch}{1}

\bibliographystyle{agsm}
\bibliography{causal.bib}
}
\end{document}